\documentclass[preprint,12pt,table]{elsarticle}

\usepackage[table]{xcolor}
\usepackage{graphicx}
\usepackage{units} %nicefrac
\usepackage{multirow}
\usepackage{amssymb}
\usepackage{pdfpages}
\usepackage[utf8]{inputenc}
\usepackage{booktabs}
\usepackage{ltablex}
\usepackage{lscape}
\usepackage{hyperref}
\usepackage{textcomp} %textopenbullet~

\usepackage{lineno}
\usepackage{amsmath}
\usepackage{dsfont} % for mathds
\usepackage{multicol}
\usepackage{relsize} % for mathlarger
\usepackage{import}
\usepackage{hyperref}
\journal{arXiv/stat.AP}

\begin{document}

\begin{frontmatter}


\title{The spatiotemporal tau statistic: a review}

%% use the cortext command for the associated  footnote;


\author[address1,address3]{Timothy M Pollington\corref{cor1}\href{https://orcid.org/0000-0002-9688-5960}{\includegraphics[height=8pt]{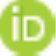}}}
\ead{timothy.pollington@gmail.com}
\cortext[cor1]{Corresponding author: MathSys CDT, University of Warwick CV4 7AL, UK}
\author[address2]{Michael J Tildesley\href{https://orcid.org/0000-0002-6875-7232}{\includegraphics[height=8pt]{figs/orcid_16x16.pdf}}} \author[address3]{T Déirdre Hollingsworth\corref{equal1}\href{https://orcid.org/0000-0001-5962-4238}{\includegraphics[height=8pt]{figs/orcid_16x16.pdf}}} \author[address4]{Lloyd AC Chapman\corref{equal1}\href{https://orcid.org/0000-0001-7727-7102}{\includegraphics[height=8pt]{figs/orcid_16x16.pdf}}}
\cortext[equal1]{Equal contributions from TDH \& LACC}

\address[address1]{MathSys CDT, University of Warwick, UK}
\address[address2]{Zeeman Institute (SBIDER), School of Life Sciences and Mathematics Institute, University of Warwick, UK}
\address[address3]{Big Data Institute, Li Ka Shing Centre for Health Information and Discovery, University of Oxford, UK}
\address[address4]{London School of Hygiene \& Tropical Medicine, UK}
\begin{abstract}

\textbf{Introduction}
The \textit{tau statistic} is a recent second-order correlation function that can assess the magnitude and range of global spatiotemporal clustering from epidemiological data containing geolocations of individual cases and, usually, disease onset times. This is the first review of its use, and the aspects of its computation and presentation that could affect inferences drawn and bias estimates of the statistic. 

\textbf{Methods}
Using Google Scholar we searched papers or preprints that cited the papers that first defined/reformed the statistic. We tabulated their key characteristics to understand the statistic's development since 2012. 

\textbf{Results}
Only half of the 16 studies found were considered to be using true tau statistics, but their inclusion in the review still provided important insights into their analysis motivations. All papers that used graphical hypothesis testing and parameter estimation used incorrect methods. There is a lack of clarity over how to choose the time-relatedness interval to relate cases and the distance band set, that are both required to calculate the statistic. Some studies demonstrated nuanced applications of the tau statistic in settings with unusual data or time relation variables, which enriched understanding of its possibilities. A gap was noticed in the estimators available to account for variable person--time at risk. 

\textbf{Discussion}
Our review comprehensively covers current uses of the tau statistic for descriptive analysis, graphical hypothesis testing, and parameter estimation of spatiotemporal clustering. We also define a new estimator of the tau statistic for disease rates. For the tau statistic there are still open questions on its implementation which we hope this review inspires others to research.
\end{abstract}


\end{frontmatter}


\newpage
%\linenumbers
\newpage

\tableofcontents
\newpage

\section{Introduction}
\label{S:Intro}

Transmission of infection is a dynamic process in time and space. Infectious diseases spread because a pathogen is transmitted by `contact' with `parent' cases (where we use `contact' in a loose sense to include transmission from the parent case(s) via a vector, airborne transmission, fomites, environmental contamination etc.). It is therefore expected that observed cases are infected by a parent case both close in time \textit{and} space. The additional distinction of a \textit{spatiotemporal} infection process is because normally any case will only be infectious for a short period relative to the study length thus leaving a temporal signal coinciding with their spatial presence. We focus on infectious diseases which require some explicit consideration of the infection process to formulate pair-relatedness variables. However, the risk factors for non-infectious diseases like poverty, environment, family/cultural traits may also follow spatiotemporal processes, like those between parent and offspring cases which we seek to measure for infectious disease. This demonstrates how non-infectious risk factors that coincide with infectious transmission may interfere with the spatiotemporal signal measured from an infectious disease dataset.

Knox defines \textit{clustering} as ``a geographically bounded group of occurrences of sufficient size and concentration to be unlikely to have occurred by chance'' \cite{Knox1989}. This could be `hotspot clustering' which is ``any area within the study region of significant elevated risk'' \cite{Lawson1999b}, or \textit{global clustering} which is a general tendency to cluster across a study. We avoid using the term `spatial/spatiotemporal dependence' as it is not clearly defined in the literature. 

Increasing availability of accurate geolocation data in recent years has enabled better understanding of this process. Being able to detect global disease clustering and assess its spatial extent can inform decisions on infectious disease control that can make better use of limited public health resources. However, standard clustering statistics in this domain often consider the spatial dimension only. (\S\ref{S:currentstats1}). 
Here we review publications using the \textit{tau statistic} \cite{Salje2012,Lessler2016}---a recent global clustering statistic for infectious disease (\S\ref{S:thetaustatistic}). In this literature review we explain its general purpose and identify sources of bias in the statistic. 
In an upcoming paper we consider how the different aspects of implementation identified here may bias the tau statistic \cite{Pollingtontech}. 

\subsection{Current statistics \& tests for global clustering}
\label{S:currentstats1}
This review focuses on the tau statistic \cite{Salje2012,Lessler2016}, but we first present other statistics for assessing disease clustering to highlight the benefits of the tau statistic to newcomers. Ward summarises spatiotemporal methods for disease data \cite{Ward2007} by those that are based on mechanistic modelling, like spatiotemporal kernel models \cite{Chapman2018}, and those based on statistical modelling, like the Matérn cluster process that describes a spatiotemporal point process; where statistics may be chosen for computational efficiency or the assumptions and sensitivities of the spatial distributions of the underlying population at risk \cite{Ward2007}. Alternatively, empirical measures can estimate global clustering of individual cases (first-order) (\S\ref{S:currentstats1}) or case pairs (second-order) (\S\ref{S:currentstats1} \& \S\ref{S:thetaustatistic}). Second-order measures are particularly appropriate for investigating the infection process \textit{between} individuals since we typically assume that infection occurs from one parent case infecting one susceptible offspring.

\subsubsection{Spatial-only or spatiotemporal tests}
\label{S:currentspatstats}
Cuzick \& Edwards' $k$-nearest neighbours test \cite{Cuzick1990},  Anderson \& Titterington's Integrated Squared Difference function \cite{Anderson1997} and Tango's C \cite{Tango1999} are tests for clustering that divide the data into cases and controls. Unfortunately they only describe clustering in the spatial dimension. These three tests assume ``two independent inhomogeneous Poisson processes with spatially-varying intensities: $m_1(\mathbf{x})$ for sampled cases and  $m_2(\mathbf{x})$ for sampled controls" \cite{Tango1999} randomly chosen from ``individuals at risk in the study region" \cite{Tango1999}.
\newline
\newline
\noindent\textit{Cuzick \& Edwards' k-nearest neighbours test} sums the number of case-case pairings within a certain range \cite{Cuzick1990}, which has similarities to the tau statistic.
\begin{equation}
\begin{split}
    T_k :=\sum_i\sum_j& a_{ij}\delta_i\delta_j,
    \textnormal{ where }\delta_i = \mathds{1}(i\textnormal{ is a case}) \textnormal{, and for locations }\mathbf{x}_j \textnormal{ of case } j\\
    a_{ij} &= \mathds{1}(\mathbf{x}_j \in k\textnormal{-nearest neighbours of }\mathbf{x}_i)
\end{split}
\end{equation}

\noindent\textit{Anderson \& Titterington's Integrated Squared Difference function ($\,\widehat{\textnormal{ISD}}$)} smooths the difference of non-parametric kernel density-estimated relative risks in cases and controls ($\hat{m}_1,\hat{m}_2$) at point $\mathbf{x}$ across 2D space $S$ \cite{Anderson1997,Tango1999}.
\begin{equation}
    \label{eq:isd}
    \widehat{\textnormal{ISD}}:=\int_{\mathbf{x}\in S}(\hat{m}_1(\mathbf{x}) - \hat{m}_2(\mathbf{x}))^2d\mathbf{x}
\end{equation}

\noindent\textit{Tango's C} imposes a parametric kernel in the $\widehat{\textnormal{ISD}}$ (Equation \ref{eq:isd}), e.g.\ a step function for hotspot clusters or exponential decay for clinal clusters \cite{Tango1999}.
\newline
\newline
\noindent\textit{Spatiotemporal K-function} initially developed for stationary point processes \cite{Diggle1995}, it has strong connections to the tau statistic as it is mentioned in appendix of the first paper to define and use the tau statistic \cite{Salje2012}, hereafter referred to as the \textit{Root} paper. Epidemiologically, its stationarity will never adequately explain a disease process and a constant intensity does not take account of population heterogeneity. 
Gabriel \& Diggle's \textit{inhomogeneous K function} extended it using a special class of inhomogeneous point processes \cite{Gabriel2009} and is available through the \texttt{stpp} R package \cite{stpp}. It requires a spatial case intensity estimate via kernel-based density estimation and a temporal estimate from time-series modelling \cite{Gabriel2009} so the calculation can be lengthy.
 
\begin{center}
    * * *
\end{center}
The tau statistic is defined in section \ref{S:thetaustatistic} and the review into its use described in section \ref{S:litreview}. We identify best practices, determine aspects of its estimation where it may be biased and make recommendations in sections \ref{S:litresults} \& \ref{S:discussion}. 
\newpage

\section{The tau statistic}
\label{S:thetaustatistic}
\subsection{A brief history}
The \textit{tau statistic}\footnote{This tau statistic is different from `Kendall's tau statistic' or `Kendall's rank correlation coefficient' which is a bivariate statistic for ordinal data \cite{bland2000introduction}.} is a non-parametric global clustering statistic which evaluates the disease frequency (risk, odds or rate) within a certain annulus around an average case and compares it to the background measure (at any distance), so is always positive \cite{Salje2012,Lessler2016}. ``It measures the tendency of case pairs to spatially cluster while implicitly accounting for their likeliness of being transmission-related temporally, making it a \textit{spatiotemporal} statistic'' \cite{Pollingtontech,Salje2012,Lessler2016}. Occasionally, space and time are swapped to measure temporal clustering instead with transmission relations based on spatial proximity.

The tau statistic was first defined and applied in 2012 by Salje et al.\ \cite{Salje2012}. In 2016 Lessler et al.\ described its context in the fields of spatial statistics and epidemiology, demonstrated robustness, formulated estimators for case-only or case \& non-case data, and reformed formulae in the \textit{Tau} paper \cite{Lessler2016}. Both these \textit{foundation} papers have inspired a steady stream of papers applying the tau statistic or similar statistics. The code to calculate both $\hat{\tau}_{\textnormal{odds}},\hat{\tau}_{\textnormal{prev}}$ estimators is available in the \texttt{IDSpatialStats} R package \cite{LesslerGiles}. Since datasets with thousands of cases can take tens of hours to construct confidence intervals for, we have re-implemented $\hat{\tau}_{\textnormal{odds}}$ \& $\hat{\tau}_{\textnormal{prev}}$ in the C language, providing a speed-up of up to 76 times \cite{Pollington2019}. For a dataset of a few hundred cases the point estimate and bootstrapped tau estimates can typically be obtained in seconds.

There are some similarities between the tau statistic and earlier statistics which focused on areas of excess risk $R$: $R(\mathbf{x}) = \lambda(\mathbf{x})/g(\mathbf{x})$ where the numerator represented the case intensity at point $\mathbf{x}\textnormal{ in space }S$ and denominator the ``background effect'' \cite{Lawson2013}. ``Some information concerning the scale of clustering can also be obtained by this method'' \cite{Lawson2010} on changing tolerance bounds to detect where clustering is strongest. The tau statistic's functional form differs as the numerator's distance band $[d_1,d_2)$ is nested within the denominator's $(0,\infty)$ as we shall see in Equations \ref{eq:tauodds}-\ref{eq:taurate}. 

\subsection{Tau statistic $\tau_\textnormal{odds}$ (odds ratio estimator)}
The distance form of the tau statistic $\tau_{\textnormal{odds}}$ is a ratio of the odds $\theta(d_1,d_2)$ of finding any case $j$ which is related to any case $i$, within a half-closed\footnote{This corrects Lessler et al.'s appendices \cite{Lessler2016} that originally used an open interval. ``It has been updated in their GitHub repository \cite{LesslerGiles} following email communication on 6 December 2018'' \cite{Pollingtontech}} annulus $[d_1,d_2)$ around case $i$, versus the odds $\theta(0,\infty)$ of finding related cases over any distance separation ($d_{ij}\geq 0$) for $N$ total cases
\begin{equation}
\label{eq:tauodds}
\begin{split}
\hat{\tau}_{\textnormal{odds}}(d_1,d_2) &:= \frac{\hat{\theta}(d_1,d_2)}{\hat{\theta}(0,\infty)}\\
\textnormal{ where }&\hat{\theta}(d_1,d_2) = \frac{\sum_{i=1}^N\sum_{j=1, j\neq i}^N\mathds{1}(z_{ij} = 1, d_1\leq d_{ij}<d_2)}{\sum_{i=1}^N\sum_{j=1, j\neq i}^N\mathds{1}(z_{ij} = 0, d_1\leq d_{ij}<d_2)},
\end{split}
\end{equation}
where $\mathds{1}$ represents the indicator function, i.e.\ is equal to 1 when its argument(s) are all true and 0 otherwise. 
It is best described as ``equivalent to ratios of multitype pair correlation functions'' \cite{Lessler2016}. Values of $\tau>1$ signify spatiotemporal clustering, $\tau = 1$ implies no clustering/inhibition and $0 < \tau < 1$ means inhibition. The odds $\hat{\theta}$ in Equation \ref{eq:tauodds} is the ratio of the number of related case pairs within $[d_1,d_2)$ to the number of unrelated case pairs. The relatedness of a case pair $z_{ij}$ is determined using temporal (close onset times $t_i,t_j$), serological (same serotypes) or genotype information (e.g.\ most recent common ancestor within a time difference of the earliest onset of the pair \cite{Salje2017}) \cite{Lessler2016}. ``Typically temporal relation is defined when case onset times are within a single serial interval of each other'' \cite{Pollingtontech}. This relatedness is a probable but not certain statement of direct transmission; still the tau statistic is able to recover a spatiotemporal signal in many of its applications. Sometimes an expanding disc is chosen so we set $d_1 = 0$, relabel $d=d_2$ and have $\tau(d)$ instead.
$\tau_{\textnormal{odds}}$ is similar to an odds ratio in that it is a `ratio of odds' yet note how the numerator's distance condition ($d_1\leq d_{ij}<d_2$) is a subset of the denominator ($\forall d_{ij}\geq 0$), whereas traditionally an odds ratio is between two mutually exclusive conditions.

\subsection{Tau statistic $\tau_\textnormal{prev}$ (relative prevalence estimator)}
With the additional data of non-case locations one can compute the prevalence $\hat{\pi}(d_1,d_2)$ of related case pairs within a certain annulus versus any case or non-case pairing, and thus the prevalence form of the tau statistic approximates a risk of onset \cite{Lessler2016}. The tau statistic then becomes the relative prevalence of related case pairs within an annulus versus at any distance from an average case $i$ (Equation \ref{eq:tauprev}). Note that $N$ now represents the number of cases and non-cases combined.
\begin{equation}
    \label{eq:tauprev}
\begin{split}
\hat{\tau}_{\textnormal{prev}}(d_1,d_2) &:= \frac{\hat{\pi}(d_1,d_2)}{\hat{\pi}(0,\infty)}\\
\textnormal{ where }&\hat{\pi}(d_1,d_2) = \frac{\sum_{i=1}^N\sum_{j=1, j\neq i}^N\mathds{1}(z_{ij} = 1, d_1\leq d_{ij}<d_2)}{\sum_{i=1}^N\sum_{j=1, j\neq i}^N\mathds{1}(d_1\leq d_{ij}<d_2)}
\end{split}
\end{equation}

\subsection{A new tau statistic $\tau_\textnormal{rate}$ (rate ratio estimator)}
\subsubsection{Motivation}
Closed populations are an unrealistic model of nature due to migration, births and deaths. For long studies it is inaccurate to consider all participants being exposed to infection risk for equal times. Epidemiologists take account of this varying time-at-risk through a rate statistic. Furthermore, for diseases that confer little immunity and so occur repeatedly in the same individual, the true burden of disease would be underestimated if only one case per person is counted. For example, in cholera epidemics in Bangladesh \cite{Alam2006}, O1 \& O139 strains co-circulate yet infection from either does not confer cross-protection \cite{Heymann2008}.

\subsubsection{The estimator}
\begin{figure}[!h]
\centering\includegraphics[width=1.0\linewidth]{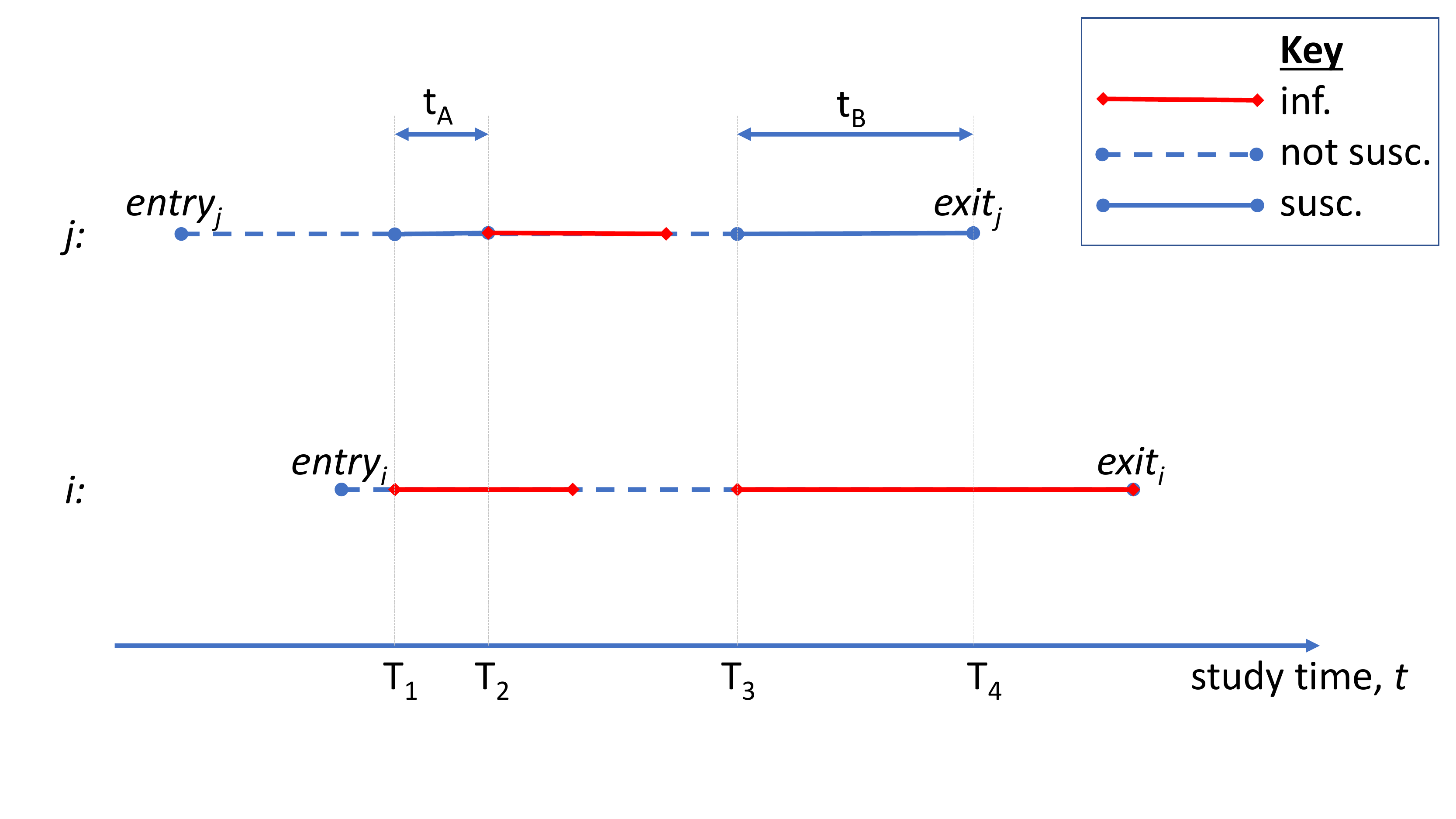}
\caption{Describing the calculation of the person-time-at-risk $(t_A+t_B)$ of individual $j$, due to $i$, by accounting for $j$'s location with respect to $i$, $j$'s susceptibility to infection (which for some diseases may depend on previous exposure) and $i$'s infectious period. 
$j$ is born and enters the study and stays within $[d_1,d_2)$ of $i$'s eventual entry. $j$ is considered to be exposed to risk of $i$ during $[T_1,T_2]$ as $j$ loses maternal immunity at $t=T_1$, thus becoming susceptible and is spatiotemporally related to $i$. At $t=T_2$, $j$ becomes infected and is no longer susceptible. One potentially-related episode pair $i\rightarrow j$ is counted. $j$ recovers but is exposed again to risk from $i$ during $[T_3,T_4]$ when $i$ becomes infectious again (in this example previous exposure does not confer protection). $j$ leaves the study, and does not experience a second episode w.r.t.\ $i$ (at least not when when enrolled on the study).}
\label{fig:rate}
\end{figure}

A rate ratio estimator of the tau statistic $\tau_{\textnormal{rate}}$ can be calculated for data consisting of cases and non-cases, with time-varying geolocations, study entry and exit times and onset and recovery times for each disease episode (Equation \ref{eq:taurate}). 
\begin{equation}
    \label{eq:taurate}
    \tau_{\textnormal{rate}}(d_1,d_2) := \frac{\lambda (d_1,d_2)}{\lambda(0,\infty)}
\end{equation}
From first principles, the incidence rate $\lambda$ is traditionally defined as the number of new events divided by the person--time-at-risk \cite{Porta2008}. For this second-order statistic we counted not the onset of a case but the transmission event across pairs. Individual $i$ is allowed to have zero to multiple disease episodes. Each single episode $l$ for individual $i$ will result in multiple probable pair episodes ($l\rightarrow m$), each of which are linked to the multiple episodes $m$ of a particular individual $j$ within $[d_1,d_2)$ of $i$ and within a certain time difference $(t_m - t_l)$. So for $n_r$ people at risk with $n$ total disease episodes during the study period, $\lambda$ is described by summing episodes in the numerator and person--time-at-risk in the denominator, unlike $\tau_{\textnormal{odds}}$ \& $\tau_{\textnormal{prev}}$ which sum cases or cases \& non-cases, respectively (Equation \ref{eq:lambda}).
\begin{equation}
    \label{eq:lambda}
    \lambda(d_1,d_2) = \frac{\sum_{l=1}^{n}\sum_{m=1,k_l\neq k_m}^{n}\mathds{1}(z_{lm} = 1, d_1\leq d_{lm}<d_2)}{\sum_{i=1}^{n_r}\sum_{j=1,j\neq i}^{n_r}\sum_{t=1}^{T_{\textnormal{end}}}\mathds{1}(Z_{ij}(t)=1,d_1\leq d_{ij}(t)<d_2)}
\end{equation}
\begin{equation*}
     \textnormal{where }Z_{ij}(t) = \mathds{1}(([\textnormal{inf.\ start}_i,\textnormal{inf.\ end}_i]\cap[\textnormal{susc.\ start}_j,\textnormal{susc.\ end}_j]\cap[t])\neq \{\phi\})
\end{equation*}
and $k_l(=i),k_m(=j)$ denote the indices of the individual to which the episodes belong; the denominator describes the total pair time at risk and is the length of time each $i,j$ pair could be potentially relatable in space and time; $[\textnormal{inf.\ start}_i,\textnormal{inf.\ end}_i]$ represents the infectious period of individual $i$ while $[\textnormal{susc.\ start}_j,\textnormal{susc.\ end}_j]$ represents $j$'s period of susceptibility, given the immunising effects of previous infection (if appropriate) and time spent in different locations $\mathbf{x}_j(t)$ relative to other $i$'s (Fig.\ \ref{fig:rate}). 

Of course for a self-immunising disease, a pair will only share one episode at most; $\tau_{\textnormal{rate}}$ is still useful in this instance as different pair times at risk still need to be accounted for. Note that the alternate calculation of $i$'s person--time-at-risk due to $j$ is not symmetric. This means $\tau_\textnormal{rate}$ will take longer to compute than $\tau_\textnormal{odds}$ or $\tau_\textnormal{prev}$ as we cannot assume transmission pairs are undirected.
A proof of concept for a real-world dataset with large migratory movements is needed to see if $\tau_{\textnormal{rate}}$ provides a substantially different estimate to $\tau_{\textnormal{prev}}$ or $\tau_{\textnormal{odds}}$ in terms of $\tau(d)$ and the range of spatiotemporal clustering $[0,D]$.

\subsection{Statistical characteristics}
Lessler et al.\ have found the following through epidemic simulations \cite{Lessler2016}:

\begin{itemize}
\item The range of clustering is consistent whether computed from the relative prevalence $\tau_{\textnormal{prev}}$ or odds ratio $\tau_{\textnormal{odds}}$ estimators.
\item Only one temporal, serotype or genotype relatedness metric is needed to infer related pairs. However more metrics will better identify true transmission pairs so that the range of clustering will less resemble the area of elevated prevalence and more the area of elevated risk, thus reducing the range of clustering and increasing the magnitude of $\tau$ in this region. 
\item In addition to the K function that estimates the clustering range, the tau statistic gives the relative magnitude of disease risk, odds or rate versus the background; this could provide informative priors for later mechanistic Bayesian modelling.
\item It is robust to population spatial heterogeneities: correctly identifies no clustering in a spatially-clustered population unlike the pair correlation function. It consistently estimates the range of clustering when only a random 1\% of cases are observed or if there is spatial observation bias e.g.\ around a surveillance outpost. This is because it is ``robust to heterogeneities in sampling probability over a study area, as the probability of sampling will similarly affect both the numerator and the denominator'' \cite{Lessler2016}. 
\item Diseases with an effective reproductive number (``the average number of people someone infected at time $t$ can infect over their infectious lifespan'' \cite{Fraser2007}) $R_e > 1$ will overestimate the clustering range while underestimating the magnitude of the $\tau$ in the true region of clustering.
\item Edge corrections are unnecessary.
\end{itemize}
\newpage

\section{Methods}
\label{S:litreview}

\subsection{Search strategy, selection \& data extraction}
We collected publicly-available works on 10 January 2019 that have cited the Root or Tau paper including articles (full text or abstract), conference abstracts, books, preprints, theses and dissertations in any language. We used Google Scholar to find articles (referred to as set B) that cite set A (either the Root paper \cite{Salje2012} or Tau paper \cite{Lessler2016}); excluding duplicates. We also looked for articles that cited set B, called set C: in case set C only referred to the closest paper of inspiration from set B rather than set A. We checked active forks from GitHub repositories of the \texttt{IDSpatialStats} package \cite{Lessler2018,LesslerGiles}. We also searched google.com for webpages and blogs about the ``tau statistic'', with disambiguation exclusions. We also announced our review to some of the previous paper authors (Salje, Lessler, Truelove \& Cummings) to inquire about any work in their research groups which was soon due for submission. We only accepted those which actively used the statistic in their analyses; mere citations to mention a previous clustering result for that particular disease were disregarded.  

The remaining works were then read fully and we contacted the papers' corresponding authors to clarify missing information; furthermore, following manuscript submission to \textit{arXiv}, we gave them a `right-to-reply' on 1 December 2019 to our commentary on their papers. This is some of the metadata extracted to summarise and find similarities, and ensure reviewing consistency:
\begin{itemize}
\item Disease
\item Format of work (article, preprint, report etc)
\item Country \& setting
\item Study type (cohort, cross-sectional, etc.)
\item Sampling method for the data
\item Calculation method of the tau statistic
\item How they presented results of the tau statistic in text \& graphics
\end{itemize}
This review is restricted to the use of the tau statistic, as consistently described in the Root \& Tau papers only. Although we found papers which claimed but did not in fact use the tau statistic, this is not a critique of their analyses. We still considered papers claiming to use the tau statistic because we wanted to review a broad spectrum of analyses based on the authors' belief that it was a tau analysis, which is still relevant to this review.
\newpage

\section{Results}
\label{S:litresults}
\subsubsection{Source information}
For works already online prior to journal publication we recorded its later journal version in the bibliography. Google Scholar found 16 papers (including the Root \& Tau papers) \cite{Salje2012,Lessler2016,Grantz2016,Grabowski2014,Succo2018,Salje2016,Bhoomiboonchoo2014,Salje2016social,Truelove2019,Azman2018,Finger2018,HoangQuoc2016,Salje2018,Rehman2018,Levy2015,Salje2017} that claimed to use the tau statistic in their analyses (Table \ref{fig:litreview}); 61 papers that mentioned the Root and 6 the Tau paper without using the statistic were ignored. There was no active code, webpages nor blogs about the statistic. All were peer-reviewed articles or reports except one recent preprint \cite{Rehman2018}; all peer-reviewed works were from respected journals with a minimum recent impact factor of 2\textperiodcentered8.
The Root paper was published in June 2012 and 15 separate works followed in 2014(\textit{2})\footnote{Italicised numbers in round brackets indicate the number of papers.}; 2015(\textit{1}); 2016(\textit{5}); 2017(\textit{1}); 2018(\textit{5}) \& 2019(\textit{1}) (Table \ref{fig:litreview}:col.\ 1). The Tau paper published in May 2016 saw 10 papers follow it. There were seven that cited both the Root \& Tau papers \cite{Truelove2019,Azman2018,Finger2018,HoangQuoc2016,Salje2018,Rehman2018,Salje2016social} and a further seven that cited the Root only \cite{Levy2015,Salje2017,Grantz2016,Grabowski2014,Succo2018,Salje2016,Bhoomiboonchoo2014}. All papers had multiple authors and always involved Salje and/or Lessler.

\subsubsection{Statistic purpose}
The main use was determining clustering and its strength but two novel alternatives were: calibrating an Approximate Bayesian Computation model by adding $\tau$ as summary statistic that captured global clustering \cite{Finger2018}; and empirically as a stopping criterion once a random labelling algorithm had reached a certain global clustering threshold \cite{Truelove2019} (Table \ref{fig:litreview}:col.\ 5).

\subsubsection{Disease spectrum \& study location}
The papers covered seven human diseases---chikungunya, cholera, HIV, influenza/influenza-like illnesses/upper respiratory illnesses, measles, pneumonia, and dengue which made the most appearances(\textit{8}) (Table \ref{fig:litreview}:col.\ 2). Analyses cover all populated continents except South America and Oceania. The tau statistic or related statistics has been used in settings where the region is a substantial landmass \cite{Truelove2019} or where there were spatial restrictions nearby or through their populations due to rivers \cite{Azman2018,Finger2018,HoangQuoc2016,Salje2012,Rehman2018,Salje2017,Salje2016,Bhoomiboonchoo2014}, walkways \cite{Levy2015} or major roads \cite{Succo2018} (Table \ref{fig:litreview}:col.\ 3).

\subsubsection{Study length, region size \& spatial/temporal resolution}
The tau statistic is most commonly used in cohort studies, which ranged from 3 months to 5 years with median 15\textperiodcentered5 months (Table \ref{fig:litreview}:col.\ 6). Two studies used cross-sectional data. 

The spatial resolution of field studies was often constrained by GPS receivers i.e.\ $\sim 10\textnormal{m}$ (Table \ref{fig:litreview}:col.\ 6). When relying on patient's reported street address to extract a geolocation, large spatial errors of 110m-1km were calculated when validated with a household visit \cite{Salje2012,HoangQuoc2016}. Furthermore cases may be aggregated at a higher spatial level because of gridded population data \cite{Finger2018} or too few cases during the study period \cite{Bhoomiboonchoo2014}. 
The temporal resolution in days, weeks or months was ultimately constrained by the reporting system. For tau papers which explicitly reported it, temporal resolution was as follows: cholera 1 day(\textit{2}); dengue 1 day(\textit{1}), 1 month(\textit{2}); measles 1 day(\textit{1}).

One paper used data with a temporal resolution similar to the length of the serial interval \cite{HoangQuoc2016} (Table \ref{tab:SI}) which is not ideal: as it could miss additional transmission pairs ($i\rightarrow j$ and $j\rightarrow k$) as conceivably within the mean 15-17 day incubation period, a case $i$ may infect $j$ and it infect a secondary case $k$, yet at monthly resolution only $i\rightarrow k$ would be observed. 

\subsubsection{Tau statistic estimators \& bootstrapping}
We consider only eight of the 16 papers that claimed to use the tau statistic actually used the form defined in the foundation papers \cite{Salje2012,Lessler2016} (Table \ref{fig:litreview}:col.\ 4).
Not all papers used the same estimators as $\tau_{\textnormal{odds}},\tau_{\textnormal{prev}}$ defined in the founds ratio estimator for case-only data (Equation \ref{eq:tauodds}) was the most common because studies typically collect the geolocation of only cases; the distance form appears in three of the 16 papers reviewed \cite{Azman2018,Finger2018,HoangQuoc2016} with the lesser-known time form in two \cite{Azman2018,Salje2018}. 

The prevalence estimator (Equation \ref{eq:tauprev}) appeared in three papers \cite{Salje2012,Grabowski2014,Salje2016social}. Despite odds ratios and risk ratios not being mathematically equivalent and some papers using the term `risk' generically for all disease measures, at low prevalences of $\sim 1$\% they are effectively equal \cite{Cummings2009}. 

The rate estimator $\tau_{\textnormal{rate}}$ we defined in \S\ref{eq:taurate} is yet to be used. We found one application of a rate-style risk ratio that varied with distance \cite{Levy2015}. The choice of a rate made sense as the epidemiological unit was respiratory illness events---something a person could have repeatedly. However they did not explicitly account for variable times at risk, presumably because they assumed that all participants stayed throughout the study.

The time form $\tau(t_1,t_2)$ swaps space and time and defines case pairs as related if they are within a specific distance band around a case. It is then evaluated across a set of time bands. Using a panel plot for different distance windows helps map out a ``dynamic risk zone'' \cite{Azman2018} akin to a simpler representation of the 2D spacetime tau colour map \cite{Salje2012}. We still consider this a tau statistic as spatiotemporal information is retained, just presented in a different way.

The number of bootstrap samples chosen had a wide range: 100(\textit{2}), 500(\textit{6}), 1000(\textit{4}), 10,000(\textit{1}) or unknown(\textit{2}). 

\subsubsection{Case definitions \& misclassification}
In most situations the case definitions were of a clinical standard beyond those typically employed for surveillance (Table \ref{fig:litreview}:col.\ 4). For diseases with a particularly fast progression from onset-to-death like cholera, people may die before reaching the hospital thus causing the data to be left-censored. There may be misclassification if the case definition of say cholera (acute watery diarrhoea at any age \cite{Heymann2008}) shares signs or symptoms with other pathogens like E.coli, shigella etc. Consideration is needed if the planned control is expected to prevent these related pathogens too and thus overestimate the potential reduction in the magnitude of tau at close distances. Finally it is unclear how the statistic would perform for an unknown disease with a broadly defined case definition e.g.\ in the initial stages of an epidemic.
  
There is potential misclassification of probably-related pairs if other contact activities occur by a different process than described by case dwelling proximity \cite{Levy2015}. For instance in Levy et al.\ (a study of respiratory illness in military recruits) considered bed location in sleeping quarters as the spatial unit that would describe the infection process whereas activities like attending a crowded mess hall could be opportunities for infection.

The example of Grabowski et al.\ \cite{Grabowski2014} challenges the assumption that using a more recent marker of infection (i.e.\ incident rather than prevalent HIV) will better identify probable transmission pairs thus leading to a stronger tau signal. The likely explanation for prevalent-incident case pairings showing higher relative risk within the household than incident-incident pairs is likely due to the low per act risk of HIV-1 infection for heterosexual vaginal sex in a developing country setting (0\textperiodcentered08\% \cite{Boily2009}) and given the relatively short study (18 months) it would have been more likely to receive reports of prevalent-incident case pairs than incident-incident.

\subsubsection{Graphical presentation}
\begin{itemize}
\item The general use of two continuous lines to represent the upper and lower parts of a series of pointwise confidence intervals is unhelpful to the untrained reader. Instead plotting each point estimate with its own confidence bands like Salje et al.\ \cite{Salje2016social} encourages the reader to consider each in turn. However since this is a common reader mistake, a default warning in the caption may be required too.
\item Most tau papers use a $\tau$-versus-distance graph (and one $\tau$-versus-time \cite{Azman2018}) to show the magnitude of $\tau$ varying with distance.
\item The convention is to plot $\tau(d_1,d_2)$ at the midpoint of the distance band like in Lessler et al.\ \cite{Lessler2016} i.e.\ $d=$\textonehalf$(d_1+d_2)$, but may be misinterpreted if not explained in the caption. The ideal default would plot the end of the distance band instead, unless the graph is used for parameter estimation purposes and then the midpoint makes sense.
\item For within-household transmission the spatial aspect of the infection process is no longer modelled as household members have no spatial freedom to move as their house is modelled as a point. It may therefore be misleading to plot a line joining $[d=0,\tau(d=0)]\leftrightarrow [d_{2^{\textnormal{nd}}},\tau(d_{2^{\textnormal{nd}}})]$ unless the first distance band includes non-zero distances i.e.\ $d\geq 0$.
\item Plotting the tau axis on a log scale can aid identification of the curve's structure. However a log scale for the distance axis may affect accurate $\hat{\tau}(d) = 0$ determination.
\item All point estimates should include envelopes unless multiple point estimates are displayed.
\item Some plot too many tau lines on the same graph \cite{Azman2018,Salje2016}. This is discouraged for more than three envelopes---aligned panel plots are an alternative.
\item The graph should cover the full extent of both bounds of the confidence interval. The axes' lines should meet at the origin so that the reader can easily read off values. The horizontal line for $\tau = 1$ is always helpful.
\item The figure caption should note the tau estimator, envelope type, number of bootstrap samples \cite{Simpson1986} and definition for time relatedness: since the graph's shape is dependent on these values. \item The Root paper (\cite{Salje2012}:Fig.\ 3) offers an advanced 2D colour plot where each pixel represents the tau estimate for a given distance and time lag. For diagnostic purposes this would be appropriate for a disease of an unknown aetiology where a diagnostic plot for initial explanatory analysis is required because the serial interval is approximate. As well as the spatiotemporal signal of primary transmission, it can reveal seasonality (through repeated regular patterns in the temporal axis) and the immunising effect of each serotype \cite{Salje2012}. However like spatiotemporal variograms the number of pairs that are separated by long spatial or temporal lags reduces, requiring caution near the plot's extremities.
\end{itemize}

\subsubsection{Distance band choice}
\label{S:choicedistbands}
The distance from an average case $i$ can be represented by a half-closed annulus with distance band $[d_1,d_2)$ or as an open disc $[d_1 = 0, d_2$). The choice depends on the purpose of analysis. An annulus will give a more precise estimate closer to some `instantaneous' $\tau$, but conversely as narrower distance bands contain fewer pairs, $\tau$ will become more variable and lead to a $\tau$-versus-distance graph that is spikier with an indiscernible trend. Alternatively an open disc conveys the cumulative risk up to a said distance $d_2$ for use by policymakers: it represents how fieldworkers operate i.e.\ up to a fixed distance from an index case $[0,d_2)$, rather than a complicated annulus shape. However open discs also smooth any intermediary spatiotemporal structure like village-to-village. Smoothing can be accentuated further by allowing distance bands to overlap as at least three papers do \cite{Lessler2016,Azman2018,Truelove2019}\footnote{we learned of this for \cite{Lessler2016} through their analysis code that they kindly shared with us}. Also as $d_2$ increases, annuli will cover more pairs so that the estimate's variance changes with distance which is detrimental to the performance of global envelope tests \cite{Baddeley2015} (a method that will soon be employed for graphical hypothesis testing in \cite{Pollingtontech}). Setting bins with equal numbers of pairs may solve this.

The tau statistic may in theory be definable at a specific single distance lag and time lag from a case to describe an instantaneous relative measure of risk, however it could never be estimated for a real data set as we only have a finite collection of points in spacetime, and apart from household transmission ($d = 0$), a given space$\circ$time lag combination is likely to exist for one pair at most. We therefore have to settle for distance bands. This thought experiment demonstrates how the existence of a true tau statistic is not known unless a future statistical proof can show asymptotic convergence to a particular limit as the number of points tend to infinity or a distance band width tends to zero. It is also telling that even if we know the transmission tree the estimate is still dependent on the distance bands we choose. Minimisation of the mean squared error ($(\hat{\tau} - \tau)^2 + \textnormal{Var}(\hat{\tau})$) is tempting but even the `true' $\tau$ is dependent on the set of distance bands. In conclusion this highlights a problem for internal validity, as for the same dataset we can arrive at a non-unique tau estimate.  

\subsubsection{Variables used to define relatedness}
`Location \& time' are the common variables(\textit{5}) used to identify probable-related transmission pairs (Tables \ref{tab:txrelation} \& \ref{fig:litreview}:col.\ 4).

\begin{table}[h]
\begin{tabular}{|c|l|c|c|}
\hline
\multicolumn{2}{|c|}{\multirow{2}{*}{}} & \multicolumn{2}{c|}{\textbf{frequency}} \\ \cline{3-4} 
\multicolumn{2}{|c|}{} & \textbf{tau studies} & \textbf{non-tau studies} \\ \hline
\multirow{4}{*}{location \newline +} & case time$^*$ & 3 & 4 \\ \cline{2-4} 
 & case time \& serotype & 4 & 0 \\ \cline{2-4} 
 & case time, serotype, MRCA$^{**}$ time & 0 & 1 \\ \cline{2-4} 
 & serostatus or none & 1 & 3 \\ \hline
\end{tabular}
\caption{Epidemiological variables used in the papers' statistics to describe transmission-relatedness of pairs. *presentation, admission or onset time of the case. **most recent common ancestor.}
\label{tab:txrelation}
\end{table}

Some authors purported use of a `tau statistic' lacked a temporal element \cite{Truelove2019} thus reducing it to a spatial statistic. Either the papers used the phi statistic $\phi$ (a related statistic concerned with spatiotemporal interaction \cite{Salje2012}) \cite{Bhoomiboonchoo2014,Grantz2016,Rehman2018}, or risk \cite{Levy2015,Salje2016,Salje2017,Truelove2019} or odds ratios \cite{Succo2018}. For instance, for $\pi(d)$ (the numerator of $\tau_{\textnormal{prev}}$) Levy et al.\ used the probability of finding sick pairs within distance $d$ out of all sick pairs, rather than the probability that pairs found within $d$ are sick, while their denominator for $\tau$ was the proportion of pairs within $d$ rather than the proportion of sick pairs with $d$ compared to all pairs. Similarly others make $\tau$ the ratio between seroconverted and all individuals \cite{Salje2016,Salje2017} or cases and non-cases \cite{Succo2018} rather than between the risk/odds of finding a case within a distance versus at all distances.

Grabowski et al.\ \cite{Grabowski2014} is a unique example of the tau statistic where no temporal, geno nor serotype information is needed to link pairs---through an implicit temporal relation. Since a prevalent case is defined as having HIV before the study, and incident cases are those detected during the 19 month study, a temporal relation between prevalent and incident cases can be formed. This may be a useful workaround if explicit onset data is unavailable for your study. 
All authors use case or virus pairs to represent the transmission chain, except Grantz et al.\ \cite{Grantz2016} who use death pairings; but this limits what can be inferred about transmission: the distribution of deaths is the convolution of the transmission process (of interest to us) with the infection-to-death process, where the latter would be confounded by local poverty and access to healthcare. However practically deaths may be the only available variable from the initial assessment of an outbreak of an unknown cause.

\subsubsection{Defining time-relatedness using serial intervals}
The aim is to represent pairs involved in primary transmission i.e.\ a single, direct transmission event between parent case $i$ and offspring case $j$ so it has been common to choose a time-relatedness interval with length equal to a single serial interval (Table \ref{fig:litreview}:col.\ 4). 
We compare the time intervals chosen against published serial intervals (Table \ref{tab:SI}). Conceptually there is sufficient reason to believe $\tau$ is sensitive to the choice of the $[t_i=T_1,t_j=T_2]$ interval, but if so then to what extent and how to find the `optimal' $[T_1,T_2]$? It is not solely about maximising specificity as for some diseases, altering $T_1,T_2$ to minimise co-primary and secondary transmission, may result in nothing left of the primary transmission peak. Alternatively a poor choice could contaminate the primary transmission signal with other indirectly-related transmission chains like coprimaries or the primary cases $k$ of $j$. Additionally we do not yet know how the effect of transmission contamination biases the true spatiotemporal signal of primary transmission. 

It is common for studies to use a particular interval without reference to the source, except Azman et al.'s cholera study \cite{Azman2018}. As a caution to future tau statistic users, the reliability of published incubation period parameters is poor e.g.\ a sample of respiratory viral infections found half did not cite the source \cite{Reich2011}. It is not just the length of the interval that is of interest but the start and endpoints ($T_1,T_2$) too---papers commonly use $T_1 = 0$ and set $T_2$ to the mean serial interval (Table \ref{tab:SI}). 

Azman et al.\ \cite{Azman2018} are nuanced in their $[T_1,T_2]$ selection for intepretative purposes. Initially they chose [0, 5d]---sensible as cholera can have an incubation period as short as a few hours \cite{Heymann2008}. However they switch to [1, 5d] to show the elevated risk in cases that they could avert i.e.\ it is unrealistic to be expected to respond to the reported onset of case $i$, to mitigate a same-day onset of $j$. 

\begin{table}[ht]
\begin{tabular}{|l|l||l|}
\hline
\textbf{Disease} & \textbf{Serial interval chosen} & \textbf{Published source} \\ \hline
Cholera & \begin{tabular}[c]{@{}l@{}}{[}0, 5d{]}(\textit{2})\\ {[}0, 4d{]}(\textit{1})\\ {[}1, 4d{]}(\textit{1})\\ {[}0, 5d{]},\ldots,{[}25d, 30d{]}(\textit{1})\end{tabular} & median 5d, range 1-11d \cite{Weil2009,Azman2016} \\ \hline
Dengue & \begin{tabular}[c]{@{}l@{}}Same month {[}0, 0mo{]}(\textit{1})\\ {[}1, 3mo{]}(\textit{1})\\ {[}3, 4-30mo{]}(\textit{1})\end{tabular} & mean 15-17d \cite{Aldstadt2012} \\ \hline
Measles & {[}0, 2wk{]}(\textit{1}) & mean 11\textperiodcentered7d \cite{Vink2014}, 14\textperiodcentered9d \cite{Cori2013} \\ \hline
\end{tabular}
\caption{Serial intervals featuring in reviewed articles (paper frequencies in round brackets) compared with values from published sources. Papers choosing variable times \cite{Salje2016social} or model-informed times \cite{Salje2018} have been excluded.}
\label{tab:SI}
\end{table}

\subsubsection{Testing spatiotemporal clustering \& estimating its range}
\label{S:determinerange}
It is common for authors to test the evidence against no spatiotemporal clustering using visual inspection of a $\tau$-versus-distance graph, which is a \textit{graphical hypothesis test}. As detailed in \cite{Pollingtontech}, all papers incorrectly estimated this range in two ways and incorrectly simultaneously established the significance of clustering: 

\begin{enumerate}[i)]
\item most construct bootstrapped estimates around the point estimate to form a \textit{central envelope} with a particular upper and lower bound according to a series of pointwise confidence intervals; they chose the endpoint of the clustering range as where the lower bound of the central envelope touched $\tau = 1$.
\item one paper \cite{Salje2012} randomly permuted the time marks $t$ across all cases (with points $(x,y,t)$) to simulate a process with no spatiotemporal clustering. An envelope was constructed about these simulations that straddled $\tau = 1$ to form a \textit{null envelope} to simulate $H_0: \tau=1$; where the point estimate touches the upper bound marks the endpoint. Again the upper and lower bounds are defined by a series of pointwise confidence intervals. 
\end{enumerate}

In \cite{Pollingtontech}, we propose corrections for the hypothesis test of no clustering $H_0: (\tau = 1)$ using global envelope tests \cite{Myllymaki2017} and estimation of the range of clustering as the horizontal set of points where the bootstrapped simulations $\hat{\underline{\tau}}^*$ intersect $\tau=1$intersection points where the bootstrapped simulation $\hat{\tau}^*(d) = 1$ (as kindly suggested by Peter Diggle in a Skype conversation on 22 October 2019). The latter also provides a measure of precision for the clustering range, unavailable under the existing methods.  

In essence the methods of the reviewed papers are incorrect as they:

\begin{itemize}
\item mix graphical hypothesis testing (which can only give a binary answer of accept/reject no spatiotemporal clustering i.e.\ $H_0:\tau=1$) with parameter estimation. Nearly all authors determine the range when the lower bound of the confidence interval touches 95\%. Azman et al.\ \cite{Azman2018} takes account of the uncertainty in the range of spatiotemporal clustering by requiring that the lower confidence bound has crossed unity over two consecutive distance bands or the median distance when bootstrap samples fall below 1\textperiodcentered2. However this is arbitrary as we do not have a theoretically-informed correction factor. 
\item Pointwise confidence intervals are common to describe the uncertainty in $\hat{\tau}$ however many authors \cite{Salje2012,Grabowski2014,Lessler2016,HoangQuoc2016,Azman2018} incorrectly use them for hypothesis testing to assert ``statistically significant'' \cite{Salje2012,Grabowski2014} results: it is incorrect to scan the graph and search at multiple points along $d$ where the bound of $\tau$ or null envelope is first crossed and then declare that as the clustering endpoint. Since multiple pointwise CIs are compared with $\tau=1$, during this inspection it amounts to a series of multiple hypothesis tests which inflates the chance a true null hypothesis is rejected (type I error).
\item Their method also cannot estimate the uncertainty of the clustering range parameter $D$ estimated.
\end{itemize}

All papers use two-sided confidence intervals. This is sensible as immunising effects could cause inhibition at close distances. Although for a well-studied disease with known localised clustering, there may be reason to choose a one-tailed test apriori.

\subsubsection{External validity}
The tau statistic has been well tested in a range of infectious diseases exploring person-to-person and vector transmitted diseases, with short to medium serial intervals and different markers of case relatedness (Table \ref{fig:litreview}:cols.\ 2 \& 4). The study settings have ranged from urban, peri-urban to rural setting at different population densities (Table \ref{fig:litreview}:col.\ 3).

The tau statistic has not yet been applied to diseases like leprosy or visceral leishmaniasis whose highly variable incubation periods would increase the uncertainty in the clustering range \cite{Heymann2008}.

\section{Discussion}
\label{S:discussion}
Clustering is an important characteristic to shed light on infection dynamics and can inform disease control or academic study. The tau statistic has been applied for this purpose to disease datasets which contain the location of cases (and possibly non-cases) and some variables to link probable transmission pairs by temporal, serological or genotypic attributes.

This review surveyed a number of papers which claimed to analyse disease clustering using the tau statistic. We only considered half of the 16 papers reviewed to truly use the tau statistic as originally defined by the foundation papers \cite{Salje2012,Lessler2016}; this was either because a temporal element was lacking or their formulae were better described as a risk/odds ratio or phi statistic. Lessler et al.'s analysis demonstrated robustness of the statistic \citet{Lessler2016}. However, we caution readers that they cannot necessarily expect these same benefits or caveats to apply for statistics outside the $\tau_{\textnormal{odds}}$ or $\tau_{\textnormal{risk}}$ definition. Despite this, this review has been richer for their inclusion through learning about the authors' analysis motivations. This review has also uncovered examples of good practice or ways in which the statistic has been redefined:

\begin{itemize}
\item defining the lower interval $T_1$ of the  time-relatedness interval $[T_1,T_2]$ to equal the expected field response time to avoid overestimating the elevated risk that may be averted 
\item outcome variable of death rather than case of disease
\item plotting a `distance lag'-vs-`time lag' 2D colour plot, where each pixel represents a tau value
\item a time form of the tau statistic $\tau(t_1,t_2)$
\item considering `pre-study' (prevalent) cases and `during study' (incident) cases as a proxy to define time-relatedness on if onset times are unavailable
\end{itemize}

Knowledge about the tau statistic has been concentrated in the medium of journal articles and further limited to papers written by authors of the foundation papers---Salje \& Lessler. Yet this statistic could be very useful to infectious disease modellers, field epidemiologists and policy makers, particularly given its implementation in the freely available \texttt{IDSpatialStats} R package. To further boost adoption, we plan to provide \texttt{R Markdown} tutorials of the tau statistic on open access training hubs like RECONlearn.org.

We hope this review has given readers an appreciation of the tau statistic with caveats on its use. Like any statistic the skilled epidemiologist should still be aware of standard concepts like case definitions to avoid misclassification. On graphing the results we have mentioned a few standard practices that can present the data objectively to avoid misleading the reader. Depending on the graphical purpose of the $\tau$-vs-distance graph, open annuli for control policy questions; or open discs to investigate fine spatiotemporal structure may be appropriate, respectively. Furthermore as a spatiotemporal statistic one must consider the data's spatial resolution, temporal resolution relative to the mean serial interval, and if the space and time variables are likely to represent the actual infection process.

\subsection{Recommendations for further quantitative research}

Following this review we recommend the following aspects of the tau statistic's implementation are further investigated to assess their bias:

\begin{itemize}
\item It is currently unclear how best to choose the distance band set to reduce both bias and variance in the tau statistic and whether equi-distant or equi-number bins should compose them.
\item If using time-relatedness to link cases then how to choose the interval $[T_1,T_2]$ given a known serial interval for the disease.
\item Differences in health status or treatment-seeking/healthcare could change the disease latent period or infectious period, respectively. Would this require a reappraisal of the time-relatedness interval over the course of the study?
\item The number of bootstrap samples ranges over 100 fold but the importance of this implementation parameter is unknown.
\item Test the new rate ratio estimator $\tau_{\textnormal{rate}}$ on a dataset containing geolocations of cases and non-cases and the times of onset and recovery of disease episodes. The chosen setting should be where individuals have variable times-at-risk of disease due to seasonal migration or staggered study entry/exit; ideally the disease would be one that confers little protection following exposure so that multiple episodes are observable. Assess differences in the estimator compared to $\tau_{\textnormal{prev}}$. Given the scarcity of existing good quality data of this kind, a study may to be prospectively designed.
\item Investigate the use of the tau statistic as a (global) spatial summary statistic for Approximate Bayesian Computation. To what extent does the tau statistic help with computation accuracy and efficiency and how should it be weighted relative to other summary statistics used by the algorithm?
\item Immunity from disease exposure had a large biasing effect on the estimation of the mean transmission distance of simulated epidemics \cite{Salje2016Tx}. It is therefore sensible to assess tau statistic performance for immunising (SIR-style)\footnote{SIR is a compartmental model describing the progression of individuals from \underline{S}usceptible, through \underline{I}nfected, to \underline{R}ecovered states of a disease. SIS diseases oscillate between S and I as individuals do not gain protection following infection.} and non-immunising (SIS-style) diseases.
\item Test the validity of the tau statistic for diseases with highly variable incubation periods
\item Is the tau statistic prone like other spatial statistics to population shift bias over time?
\item Although it has been shown to robustly extract a clustering signal for a 1\% fraction of a simulated dataset \cite{Lessler2016}, what is the minimum number of cases for the tau statistic to perform reliably?
\item What is the theoretical formulation of the tau statistic and can mathematical analysis of its statistical properties bring us new insights? 
\item Considering that the tau statistic uses symmetric shapes like discs or annuli that assume an isotropic disease process, how would anisotropies in transmission e.g.\ along road networks, land relief or wind affect tau estimates?
\item Impact of spatial aggregation of misspecification versus actual infection location and seasonality, on the tau statistic.
\end{itemize}
Some of the above aspects will be covered in an upcoming quantitative study applied to a measles dataset \cite{Pollingtontech}.

\newpage

\begin{center}
    * * *
\end{center}
Control programmes have already been informed by the tau statistic so applying improvements on its implementation and further research will safeguard future health decisions based on it.

\section{Acknowledgements \& funding sources}
\label{S:ack}
We kindly thank: 
\begin{itemize}
    \item Lessler \& Salje who openly answered questions on their work. We would like to thank Truelove who replied to our questions by email.
    \item Peter Diggle for explaining proper methods for graphical hypothesis testing and methods for estimating the clustering range.
\end{itemize}

TMP, LACC \& TDH gratefully acknowledge funding of the NTD Modelling Consortium by the Bill \& Melinda Gates Foundation (BMGF) (grant \textnumero~OPP1184344), and LACC acknowledges funding of the SPEAK India consortium by BMGF (grant \textnumero~OPP1183986). Views, opinions, assumptions or any other information set out in this article should not be attributed to BMGF or any person connected with them. 

TMP's PhD was supported by the Engineering \& Physical Sciences Research Council, Medical Research Council and University of Warwick (grant \textnumero~EP/L015374/1). TMP would like to thank Big Data Institute for hosting him during this review under the supervision of Déirdre Hollingsworth.

All funders had no role in the study design, collection, analysis, interpretation of data, writing of the report, or decision to submit the manuscript for publication.

\section{Competing interests}
All authors declare no competing interests.

\section{Contributions: CRediT statement}
\noindent \textbf{TMP:} Conceptualisation, Methodology, Validation, Formal analysis, Investigation, Data curation, Writing - original draft \& editing, Visualisation \textbf{MJT:} Conceptualisation, Writing - review \& editing, Supervision \textbf{PJD:} Methodology (see \S\ref{S:ack}) \textbf{TDH:} Conceptualisation, Writing - review \& editing, Supervision, Funding acquisition \textbf{LACC:} Conceptualisation, Writing - review \& editing, Supervision.

\section{Copyright}
This article is licensed under the Creative Commons Attribution- \newline NonCommercial-NoDerivatives Works 4·0 International Licence (CCBY-NC-ND4\textperiodcentered0). Anyone can copy and distribute this article unchanged and unedited but only for non-commercial purposes, provided the user gives credit by providing this article's DOI and a link to the licence (creativecommons.org/ licences/by-nc-nd/4.0). The use of this material by others does not imply endorsement by the authors.
\newpage
\section{Figures}
\newpage

\subsection{Key characteristics of the reviewed papers}
\refstepcounter{table}\label{fig:litreview}
\begin{landscape}
{\rowcolors{5}{green!25!yellow!25}{green!15!yellow!15}
\begin{tabularx}{\linewidth}{|>{\hsize=1.0\hsize}X|>{\hsize=1.0\hsize}X|>{\hsize=0.6\hsize}X|>{\hsize=1.8\hsize}X|>{\hsize=1.4\hsize}X|>{\hsize=0.6\hsize}X|>{\hsize=0.6\hsize}X|>{\hsize=1.0\hsize}X|}
\toprule
\scriptsize\textbf{Reviewed\newline papers. Root$^R$/\newline Tau$^T$ paper cited?} & \scriptsize\textbf{Human\newline disease \& case defn.} & \scriptsize\textbf{Location rural$^r$/ urban$^u$} & \scriptsize\textbf{Statistic(\textbullet~)\newline \&\newline epidemiological unit(\textopenbullet~)} & \scriptsize\textbf{Purpose \&\newline stated findings} & \scriptsize\textbf{Study type\footnote{C = prospective cohort; RC = retrospective cohort; S = simulation; TS = time series; XS = cross-sectional} \& scale} & \scriptsize\textbf{\textnumero~ cases, events, people or deaths} & \scriptsize\textbf{Sampling method}\\
\midrule
\endhead

\footnotesize\textbf{ROOT}\newline\textbf{PAPER}\newline Salje\newline et al.\ \cite{Salje2012}\newline \tiny 2012 & \tiny Dengue\newline RT-PCR\footnote{Reverse-Transcription Polymerase Chain Reaction} (incl.\ serology) & \tiny Bangkok$^u$, TH & \tiny\textbullet~$\tau$(prevalence, distance) (case only)\footnote{Tau statistic, see \S\ref{S:thetaustatistic} for detailed information on estimators}\newline \textbullet~$\phi$(distance \& time)\footnote{Phi statistic measures spatiotemporal interaction \cite{Salje2012}}\newline\textopenbullet~ case with serotype, admission date, address, time $(t_j-t_i\leq$[1-3mo], or $\in$[3,4-30mo]$\Rightarrow z_{ij}=1$) & \tiny Spatial clustering of same-serotype cases within 1 km.\ & \tiny TS\newline 5yr\newline $\sim$1,569km$^2$ & \tiny 1,912 geocoded & \tiny hospital,\newline children\\
\footnotesize Grabowski\newline et al.$^R$ \cite{Grabowski2014}\newline \tiny 2014 & \tiny HIV\footnote{Human Immunodeficiency Virus}\newline confirmed by serology/western blot & \tiny Rakai district$^r$, UG & \tiny\textbullet~$\tau$(prevalence, distance)\newline \textopenbullet~ case/non-case pair, hhld GPS\footnote{Global Positioning System}, serostatus (every 12-18mo) & \tiny Spatial clustering of seropositive individuals from the hhld.\ level up to 250m but not at the community level & \tiny C\newline 19mo\newline $\sim$3,352km$^2$ & \tiny 14,594 people (70\% of censused popn.), 8,899 hhlds.\ 8,156/8,899 geocoded hhlds., 12.2\% HIV seroprevalence and incidence 1.2/100pyrs & \tiny community,\newline 15-49yrs\\
\footnotesize Bhoomi-\newline boonchoo\newline et al.$^R$ \cite{Bhoomiboonchoo2014}\newline \tiny 2014 & \tiny Dengue\newline confirmed by RT-PCR and IgM/IgG\footnote{Immunoglobulin M \& G antibodies} serology & \tiny Kamphaeng Phet province$^r$, TH & \tiny \textbullet~$\phi$(distance)\newline\textopenbullet~ cases, village-level GPS, time $(t_j - t_i\leq$30d$\Rightarrow z_{ij}=1)$ & \tiny Spatiotemporal clustering of cases within 1 mo and living in the same village.\ & \tiny TS\newline 14yr\newline $\sim$8,608km$^2$ & \tiny 4,768 (93\% of all cases) & \tiny hospital,\newline from villages with $\geq$40 cases\\
\footnotesize Levy\newline et al.$^R$ \cite{Levy2015}\newline \tiny 2015 & \tiny URI/ILI\footnote{Upper Respiratory Illness or Influenza-Like Illness}/\newline influenza\newline confirmed by influenza RT-PCR and multiplex PCR & \tiny Military barracks$^r$, TH & \tiny\textbullet~ risk ratio(events, distance)\footnote{Reported by authors as a tau statistic\label{fn:reportedtau}}\newline\textopenbullet~ case events, bed location, presentation time $(t_j - t_i\leq$1w$\Rightarrow z_{ij} = 1)$ & \tiny Non-significant clustering of cases up to 5m.\ & \tiny C\newline 11w\newline 1 sleeping quarter & \tiny 77 ILI/URI events, 122 recruits & \tiny 20-31yr male recruits. Pre-existing TB or immunosuppression excl.\\
\footnotesize Salje\newline et al.$^R$ \cite{Salje2016}\newline \tiny 2016 & \tiny Chikungunya\newline confirmed by febrile + RT-PCR & \tiny Cebu City$^u$, PH & \tiny\textbullet~ risk ratio(fixed distance window)$^{\ref{fn:reportedtau}}$\newline\textopenbullet~ seroconversion event(DENV\footnote{Dengue Virus} 1-4)(12mo apart),hhld.\ location & \tiny Spatial dependence of seroconversion$\leq$230m---rationale for focal interventions.\ & \tiny C\newline 1yr\newline $\sim$315km$^2$ & \tiny $\sim$106 seroconversions of 851 people & \tiny community, randomly sampled, $\geq$6mo age, only one selected per hhld.\\
\footnotesize\textbf{TAU}\newline\textbf{PAPER}\newline Lessler\newline et al.$^R$ \cite{Lessler2016}\newline \tiny 2016 & \tiny Dengue, HIV, measles & \tiny Data re-use \cite{Salje2012,Grabowski2014} \& Hagelloch$^r$, DE & \tiny \ldots\footnote{``\ldots/'' = Re-use of data mentioned elsewhere in table, see disease or location featured in the second or third columns of this row.}/\ldots/\newline\textbullet~$\tau$(prevalence, distance)\newline\textopenbullet~ onset date $(t_j - t_i\leq$2w$\Rightarrow z_{ij} = 1)$ & \tiny Reformed the use of $\tau$ w.r.t. formulae and use of case and non-case data.\ & \tiny \ldots/\ldots/\newline 3mo\newline $\sim$0\textperiodcentered06km$^2$ & \tiny \ldots/\ldots/\newline 188 & \tiny \ldots/\ldots/\newline community, children from case homes\\
\footnotesize Salje\newline et al.$^{RT}$ \cite{Salje2016social}\newline \tiny 2016 & \tiny Chikungunya\newline $\sim$48\% confirmed by IgM serology & \tiny Palpara$^r$, BG & \tiny\textbullet~ $\tau$(prevalence, distance)\newline\textopenbullet~ case/non-case pair, onset date (variable generation time, mean 14d), hhld GPS & \tiny Used to test the sensitivity of global clustering by different transmission kernel sizes of a simulated epidemic.\ & \tiny XS\newline 6mo\newline $\sim$0\textperiodcentered6km$^2$ & \tiny 1,933 individuals, 460 hhlds, 175 confirmed & \tiny community, every hhld in outbreak village\\
\footnotesize Grantz\newline et al.$^R$ \cite{Grantz2016}\newline \tiny 2016 & \tiny Influenza/\newline pneumonia\newline reported by Chicago D.o.H.\ & \tiny Chicago$^u$, US & \tiny\textbullet~ $\phi$(distance)\newline\textopenbullet~ case death pair, death date ($t_j - t_i\leq$1w$\Rightarrow z_{ij} = 1$) & \tiny Spatial clustering of mortality at the census-tract level.\ & \tiny TS\newline 7w\newline $\sim$ 606km$^2$ & \tiny 7,971 deaths & \tiny community, routine data\\
\footnotesize Hoang Quoc\newline et al.$^{RT}$ \cite{HoangQuoc2016}\newline \tiny 2016 & \tiny Dengue\newline confirmed by RT-PCR & \tiny Ho Chi Minh City$^u$, VN \& \cite{Salje2012}& \tiny\textbullet~ $\tau$(odds, distance)\newline\textopenbullet~ case pair, serology(DENV1-4), address, admission date$(t_j-t_i = 0)$mo$\Rightarrow z_{ij} = 1$), & \tiny Small-scale spatial clustering of cases $<$500m & \tiny C\newline 4yr\newline $\sim$2061km$^2$/\newline\ldots & \tiny 1,444 with serology \& geolocated/\ldots & \tiny hospital, imprecise geolocations dropped\\
\footnotesize Salje\newline et al.$^R$ \cite{Salje2017}\newline \tiny 2017 & \tiny Dengue\newline confirmed by RT-PCR or IgM/IgG serology & \tiny TH$^{ru}$ & \tiny\textbullet~ risk ratio(prevalence, distance)\footnote{The authors also analysed the spatial relationship of proportions of case pairs falling ill within 6 months and coming from the same transmission chain at different distances. However as a proportion ranges between 0 and 1 we have not included it here as it is not comparable with the positive real $\tau$.}\newline\textopenbullet~ case pair or virus pair, admission date($t_j-t_i\leq$6mo$\Rightarrow z_{ij} = 1$), serotype(DENV1-4), hhld.\ GPS, MRCA\footnote{Most Recent Common Ancestor} date ($g_j - g_i\leq$6mo, or $\in$[6mo,2yr), [5,10yr) from sequencing data) & \tiny Virus pair spatiotemporal clustering $\leq$5km \& 6mo of MRCA. & \tiny RC\newline 16yr\newline$\sim$513,120km$^2$ & \tiny 17,931\newline(=640+17,291) & \tiny hospital,\newline children or young teenagers where serotype is known\\
\footnotesize Finger\newline et al.$^{RT}$ \cite{Finger2018}\newline \tiny 2018 & \tiny Cholera\newline acute watery diarrhoea + any age & \tiny N'Djamena$^u$, TD & \tiny\textbullet~ $\tau$(odds, distance). Distance windows were constrained by ``spatial discretisation of the model domain''.\newline\textopenbullet~ case pair, hhld.\ GPS, onset date($t_j - t_i\leq$5d$\Rightarrow z_{ij} = 1$ & \tiny $\tau$ calibrated a simulation model (in equal parts with a spatially explicit individual-based stochastic model) to test different intervention scenarios. & \tiny TS\newline 7mo\newline166km$^2$ & \tiny 1,585 geolocated (of 4,352) & \tiny hospital,\newline $\sim\nicefrac{1}{2}$ cases geolocated (confirmed by home visit)\\
\footnotesize Salje\newline et al.$^{RT}$ \cite{Salje2018}\newline \tiny 2018 & \tiny Dengue\newline virus isolation + serological evidence & \tiny Kamphaeng Phet province$^r$, TH & \tiny\textbullet~ $\tau$(odds, time)\newline\textbullet~ odds ratio(place, fixed time windows)$^{\ref{fn:reportedtau}}$\newline\textopenbullet~ case pair, serotype(DENV1-4), school, augmented model infection time (assume symptomatics' median incubation period - 7d; undetecteds' infection-to-titre rise = 11d) & \tiny Model diagnostic on inferred undetected subclinical infections---augmented infections shared the temporal clustering (specific to serotype and place) as symptomatic infections. & \tiny C\newline 5yr\newline $\sim$98km$^2$ & \tiny 3,451 with fever symptoms & \tiny school\newline 8-11yr age, blood sampled every 3mo, excl.\ if migration plans within 12mo or thalassaemia.\\
\footnotesize Succo\newline et al.$^R$ \cite{Succo2018}\newline \tiny 2018 & \tiny Dengue\newline anti-DENV IgM and IgG +ve + febrile + body temp $\geq$38$^o$C + not another condition & \tiny Nîmes$^u$, FR & \tiny\textbullet~ odds ratio (fixed distance window)\footnote{Reported as a relative risk in the main text, but as an odds ratio in the Supplementary material}\newline\textopenbullet~ case/non-case pair, hhld.\ GPS, hhld.\ ID (to differentiate same bldg.\ but different hhld.) & \tiny Spatial clustering of case vs non-case pairs detected at the hhld.\ level but no further. & \tiny XS\newline 15d\newline $\sim$0\textperiodcentered6km$^2$ & \tiny 1431 people, 512 hhlds, prev.\ 0.4\% & \tiny community, residing $\geq$4mo, $\geq$2yr age\\
\footnotesize Rehman\newline et al.$^{RT}$ \cite{Rehman2018}\newline \tiny 2018 & \tiny Dengue\newline confirmed case & \tiny Rawalpindi$^u$ \& Lahore$^u$, PK & \tiny\textbullet~ $\phi$(distance \& time)$^{\ref{fn:reportedtau}}$\newline\textopenbullet~ case, hhld GPS, onset date($t_j-t_i\leq$30d$\Rightarrow z_{ij} = 1$ & \tiny $\phi$ statistic compares interaction of cases in a matched intervention/control study design. & \tiny TS\newline 4 \& 6yr\newline 259km$^2$ \& 1,772km$^2$,  & \tiny 7,890 \& 2,998 & \tiny community and hospital\\
\footnotesize Azman\newline et al.$^{RT}$ \cite{Azman2018}\newline \tiny 2018 & \tiny Cholera\newline acute watery diarrhoea + any age & \tiny \cite{Finger2018} \& Kalemie$^u$, CD & \tiny\textbullet~ $\tau$(odds, distance)\newline\textbullet~ $\tau$(odds, time)\newline\textopenbullet~ case, hhld GPS, presentation date($t_j-t_i\in$[0,4d],[1,4d],[0,5d],\ldots,[25d,30d]$\Rightarrow z_{ij} = 1$ & \tiny Rationale for targeted intervention:$\leq$100m,$\leq$1w of index case presenting & \tiny \ldots/\newline TS\newline 12mo\newline $\sim$64km$^2$ & \tiny 1,692/4,359 \& 1,077/1,146 (geolocated/all) & \tiny \ldots/\newline hospital,\newline all cases geolocated\\
\footnotesize Truelove\newline et al.$^{RT}$ \cite{Truelove2019}\newline \tiny 2019 & \tiny Measles & \tiny TZ$^{ru}$ & \tiny\textbullet~ risk ratio(prevalence of vacc.\ status, distance), sample-weighted for clusters$^{\ref{fn:reportedtau}}$\newline\textopenbullet~ time-relatedness is swapped for vacc.\ status\newline unvacc.\ proportions of DHS\footnote{Demographic Health Survey} clusters, DHS cluster GPS, cluster sampling weights, numbers per cluster & \tiny Calibration tool to produce a synthetic population with a clustering of unvacc.\ that matched the empirical value from DHS surveys. & \tiny S\newline ?yr\newline 900km$^2$ & \tiny 100,000 individuals & \tiny community, residences randomly distributed in 30x30km$^2$, vacc.\ status clustered by random swapping algorithm until empirical $\tau$ reached.\\
\bottomrule
\end{tabularx}
}
\end{landscape}
\newpage



%% References



%% New version of the num-names style
\bibliographystyle{elsarticle-num-names}
\bibliography{sample.bib}

\begin{thebibliography}{49}
\expandafter\ifx\csname natexlab\endcsname\relax\def\natexlab#1{#1}\fi
\providecommand{\url}[1]{\texttt{#1}}
\providecommand{\href}[2]{#2}
\providecommand{\path}[1]{#1}
\providecommand{\DOIprefix}{doi:}
\providecommand{\ArXivprefix}{arXiv:}
\providecommand{\URLprefix}{URL: }
\providecommand{\Pubmedprefix}{pmid:}
\providecommand{\doi}[1]{\href{http://dx.doi.org/#1}{\path{#1}}}
\providecommand{\Pubmed}[1]{\href{pmid:#1}{\path{#1}}}
\providecommand{\bibinfo}[2]{#2}
\ifx\xfnm\relax \def\xfnm[#1]{\unskip,\space#1}\fi
%Type = Inproceedings
\bibitem[{Knox(1989)}]{Knox1989}
\bibinfo{author}{E.~Knox},
\newblock \bibinfo{title}{{Detection of clusters}},
\newblock in: \bibinfo{editor}{P.~Elliot} (Ed.),
  \bibinfo{booktitle}{Methodology of enquiries into disease clustering},
  \bibinfo{publisher}{Small Area Health Statistics Unit, LSHTM},
  \bibinfo{address}{London}, \bibinfo{year}{1989}, pp. \bibinfo{pages}{17--20}.
%Type = Incollection
\bibitem[{Lawson and Kulldorff(1999)}]{Lawson1999b}
\bibinfo{author}{A.~B. Lawson}, \bibinfo{author}{M.~Kulldorff},
\newblock \bibinfo{title}{{A Review of Cluster Detection Methods}},
\newblock in: \bibinfo{editor}{A.~Lawson}, \bibinfo{editor}{A.~Biggeri},
  \bibinfo{editor}{D.~B{\"{o}}hning}, \bibinfo{editor}{L.~Emmanuel},
  \bibinfo{editor}{J.-F. Viel}, \bibinfo{editor}{R.~Bertollini} (Eds.),
  \bibinfo{booktitle}{Disease Mapping and Risk Assessment for Public Health},
  \bibinfo{edition}{first} ed., \bibinfo{publisher}{John Wiley {\&} Sons Ltd.},
  \bibinfo{address}{Chichester}, \bibinfo{year}{1999}, pp.
  \bibinfo{pages}{99--110}.
%Type = Article
\bibitem[{Salje et~al.(2012)Salje, Lessler, Endy, Curriero, Gibbons, Nisalak,
  Nimmannitya, Kalayanarooj, Jarman, Thomas, Burke, and Cummings}]{Salje2012}
\bibinfo{author}{H.~Salje}, \bibinfo{author}{J.~Lessler},
  \bibinfo{author}{T.~P. Endy}, \bibinfo{author}{F.~C. Curriero},
  \bibinfo{author}{R.~V. Gibbons}, \bibinfo{author}{A.~Nisalak},
  \bibinfo{author}{S.~Nimmannitya}, \bibinfo{author}{S.~Kalayanarooj},
  \bibinfo{author}{R.~G. Jarman}, \bibinfo{author}{S.~J. Thomas},
  \bibinfo{author}{D.~S. Burke}, \bibinfo{author}{D.~A.~T. Cummings},
\newblock \bibinfo{title}{{Revealing the microscale spatial signature of dengue
  transmission and immunity in an urban population}},
\newblock \bibinfo{journal}{PNAS} \bibinfo{volume}{109} (\bibinfo{year}{2012})
  \bibinfo{pages}{9535--9538}. \DOIprefix\doi{10.1073/pnas.1120621109}.
%Type = Article
\bibitem[{Lessler et~al.(2016)Lessler, Salje, Grabowski, and
  Cummings}]{Lessler2016}
\bibinfo{author}{J.~Lessler}, \bibinfo{author}{H.~Salje},
  \bibinfo{author}{M.~K. Grabowski}, \bibinfo{author}{D.~A.~T. Cummings},
\newblock \bibinfo{title}{{Measuring Spatial Dependence for Infectious Disease
  Epidemiology}},
\newblock \bibinfo{journal}{PLoS ONE} \bibinfo{volume}{11}
  (\bibinfo{year}{2016}) \bibinfo{pages}{1--13}.
  \DOIprefix\doi{10.1371/journal.pone.0155249}.
%Type = Article
\bibitem[{Pollington et~al.(2019)Pollington, Tildesley, Hollingsworth, and
  Chapman}]{Pollingtontech}
\bibinfo{author}{T.~M. Pollington}, \bibinfo{author}{M.~J. Tildesley},
  \bibinfo{author}{T.~D. Hollingsworth}, \bibinfo{author}{L.~A.~C. Chapman},
\newblock \bibinfo{title}{{Measuring spatiotemporal disease clustering with the
  tau statistic}}  (\bibinfo{year}{2019}). \URLprefix
  \url{http://arxiv.org/abs/1911.08022}.
  \href{http://arxiv.org/abs/1911.08022}{{\tt arXiv:1911.08022}}.
%Type = Article
\bibitem[{Ward(2007)}]{Ward2007}
\bibinfo{author}{M.~P. Ward},
\newblock \bibinfo{title}{{Spatio-temporal analysis of infectious disease
  outbreaks in veterinary medicine: clusters, hotspots and foci.}},
\newblock \bibinfo{journal}{Vet. Ital.} \bibinfo{volume}{43}
  (\bibinfo{year}{2007}) \bibinfo{pages}{559--70}. \URLprefix
  \url{https://www.ncbi.nlm.nih.gov/pubmed/20422535}.
%Type = Article
\bibitem[{Chapman et~al.(2018)Chapman, Jewell, Spencer, Pellis, Datta,
  Chowdhury, Bern, Medley, and Hollingsworth}]{Chapman2018}
\bibinfo{author}{L.~A. Chapman}, \bibinfo{author}{C.~P. Jewell},
  \bibinfo{author}{S.~E. Spencer}, \bibinfo{author}{L.~Pellis},
  \bibinfo{author}{S.~Datta}, \bibinfo{author}{R.~Chowdhury},
  \bibinfo{author}{C.~Bern}, \bibinfo{author}{G.~F. Medley},
  \bibinfo{author}{T.~D. Hollingsworth},
\newblock \bibinfo{title}{{The role of case proximity in transmission of
  visceral leishmaniasis in a highly endemic village in Bangladesh}},
\newblock \bibinfo{journal}{PLoS Neglected Trop. Dis.} \bibinfo{volume}{12}
  (\bibinfo{year}{2018}) \bibinfo{pages}{1--29}.
  \DOIprefix\doi{10.1371/journal.pntd.0006453}.
%Type = Article
\bibitem[{Cuzick and Edwards(1990)}]{Cuzick1990}
\bibinfo{author}{J.~Cuzick}, \bibinfo{author}{R.~Edwards},
\newblock \bibinfo{title}{{Spatial Clustering for Inhomogeneous Populations}},
\newblock \bibinfo{journal}{J. Royal Stat. Soc. Ser. B} \bibinfo{volume}{52}
  (\bibinfo{year}{1990}) \bibinfo{pages}{73--104}.
  \DOIprefix\doi{10.1111/j.2517-6161.1990.tb01773.x}.
%Type = Article
\bibitem[{Anderson and Titterington(1997)}]{Anderson1997}
\bibinfo{author}{N.~H. Anderson}, \bibinfo{author}{D.~M. Titterington},
\newblock \bibinfo{title}{{Some Methods for Investigating Spatial Clustering,
  with Epidemiological Applications}},
\newblock \bibinfo{journal}{J. Royal Stat. Soc. Ser. A} \bibinfo{volume}{160}
  (\bibinfo{year}{1997}) \bibinfo{pages}{87--105}.
  \DOIprefix\doi{10.1111/1467-985X.00047}.
%Type = Incollection
\bibitem[{Tango(1999)}]{Tango1999}
\bibinfo{author}{T.~Tango},
\newblock \bibinfo{title}{{Comparison of General Tests for Spatial
  Clustering}},
\newblock in: \bibinfo{editor}{A.~Lawson}, \bibinfo{editor}{A.~Biggeri},
  \bibinfo{editor}{D.~B{\"{o}}hning}, \bibinfo{editor}{L.~Emmanuel},
  \bibinfo{editor}{J.-F. Viel}, \bibinfo{editor}{R.~Bertollini} (Eds.),
  \bibinfo{booktitle}{Disease Mapping and Risk Assessment for Public Health},
  \bibinfo{edition}{first} ed., \bibinfo{publisher}{John Wiley {\&} Sons Ltd.},
  \bibinfo{address}{Chichester}, \bibinfo{year}{1999}, pp.
  \bibinfo{pages}{111--117}.
%Type = Article
\bibitem[{Diggle et~al.(1995)Diggle, Chetwynd, Morris, and
  H{\"{a}}ggkvist}]{Diggle1995}
\bibinfo{author}{P.~J. Diggle}, \bibinfo{author}{A.~G. Chetwynd},
  \bibinfo{author}{S.~E. Morris}, \bibinfo{author}{R.~H{\"{a}}ggkvist},
\newblock \bibinfo{title}{{Second-order analysis of space-time clustering}},
\newblock \bibinfo{journal}{Stat. Methods Med. Res.} \bibinfo{volume}{4}
  (\bibinfo{year}{1995}) \bibinfo{pages}{124--136}.
  \DOIprefix\doi{10.1177/096228029500400203}.
%Type = Article
\bibitem[{Gabriel and Diggle(2009)}]{Gabriel2009}
\bibinfo{author}{E.~Gabriel}, \bibinfo{author}{P.~J. Diggle},
\newblock \bibinfo{title}{{Second-order analysis of inhomogeneous
  spatio-temporal point process data}},
\newblock \bibinfo{journal}{Stat. Neerlandica} \bibinfo{volume}{63}
  (\bibinfo{year}{2009}) \bibinfo{pages}{43--51}.
  \DOIprefix\doi{10.1111/j.1467-9574.2008.00407.x}.
%Type = Manual
\bibitem[{Gabriel et~al.(2018)Gabriel, Diggle, Rowlingson, and
  Rodriguez-Cortes}]{stpp}
\bibinfo{author}{E.~Gabriel}, \bibinfo{author}{P.~J. Diggle},
  \bibinfo{author}{B.~Rowlingson}, \bibinfo{author}{F.~J. Rodriguez-Cortes},
  \bibinfo{title}{stpp v2\textperiodcentered0-3: Space-Time Point Pattern
  Simulation, Visualisation and Analysis}, \bibinfo{year}{2018}. \URLprefix
  \url{https://CRAN.R-project.org/package=stpp}.
%Type = Book
\bibitem[{Bland(2000)}]{bland2000introduction}
\bibinfo{author}{J.~M. Bland}, \bibinfo{title}{{An Introduction to Medical
  Statistics}}, Oxford medical publications, \bibinfo{edition}{third} ed.,
  \bibinfo{publisher}{OUP Oxford}, \bibinfo{address}{New York, USA},
  \bibinfo{year}{2000}. \URLprefix
  \url{https://books.google.co.uk/books?id=J-F6mwEACAAJ}.
%Type = Misc
\bibitem[{Lessler and Giles(2018)}]{LesslerGiles}
\bibinfo{author}{J.~Lessler}, \bibinfo{author}{J.~Giles},
  \bibinfo{title}{{IDSpatialStats R package development version}},
  \bibinfo{year}{2018}. \URLprefix
  \url{https://github.com/HopkinsIDD/IDSpatialStats}.
%Type = Misc
\bibitem[{Pollington(2019)}]{Pollington2019}
\bibinfo{author}{T.~M. Pollington}, \bibinfo{title}{{Tau statistic speedup}},
  \bibinfo{year}{2019}. \URLprefix
  \url{https://github.com/t-pollington/tau-statistic-speedup}.
  \DOIprefix\doi{10.5281/zenodo.3460744}.
%Type = Incollection
\bibitem[{Lawson(2013)}]{Lawson2013}
\bibinfo{author}{A.~B. Lawson},
\newblock \bibinfo{title}{{Small Scale: Putative Sources of Hazard}},
\newblock in: \bibinfo{editor}{A.~B. Lawson} (Ed.),
  \bibinfo{booktitle}{Statistical Methods in Spatial Epidemiology},
  \bibinfo{edition}{second} ed., \bibinfo{publisher}{John Wiley {\&} Sons
  Ltd.}, \bibinfo{year}{2013}, pp. \bibinfo{pages}{143--187}.
  \DOIprefix\doi{10.1002/9780470035771.ch7}.
%Type = Incollection
\bibitem[{Lawson(2006)}]{Lawson2010}
\bibinfo{author}{A.~B. Lawson},
\newblock \bibinfo{title}{{Small Scale : Disease Clustering}},
\newblock in: \bibinfo{editor}{A.~B. Lawson} (Ed.),
  \bibinfo{booktitle}{Statistical Methods in Medical Research},
  \bibinfo{edition}{second} ed., \bibinfo{publisher}{John Wiley {\&} Sons
  Ltd.}, \bibinfo{year}{2006}, pp. \bibinfo{pages}{111--141}.
  \DOIprefix\doi{10.1002/9780470035771.ch6}.
%Type = Article
\bibitem[{Salje et~al.(2017)Salje, Lessler, Berry, Melendrez, Endy,
  Kalayanarooj, A-Nuegoonpipat, Chanama, Sangkijporn, Klungthong,
  Thaisomboonsuk, Nisalak, Gibbons, Iamsirithaworn, Macareo, Yoon, Sangarsang,
  Jarman, and Cummings}]{Salje2017}
\bibinfo{author}{H.~Salje}, \bibinfo{author}{J.~Lessler},
  \bibinfo{author}{I.~M. Berry}, \bibinfo{author}{M.~C. Melendrez},
  \bibinfo{author}{T.~Endy}, \bibinfo{author}{S.~Kalayanarooj},
  \bibinfo{author}{A.~A-Nuegoonpipat}, \bibinfo{author}{S.~Chanama},
  \bibinfo{author}{S.~Sangkijporn}, \bibinfo{author}{C.~Klungthong},
  \bibinfo{author}{B.~Thaisomboonsuk}, \bibinfo{author}{A.~Nisalak},
  \bibinfo{author}{R.~V. Gibbons}, \bibinfo{author}{S.~Iamsirithaworn},
  \bibinfo{author}{L.~R. Macareo}, \bibinfo{author}{I.-K. Yoon},
  \bibinfo{author}{A.~Sangarsang}, \bibinfo{author}{R.~G. Jarman},
  \bibinfo{author}{D.~A. Cummings},
\newblock \bibinfo{title}{{Dengue diversity across spatial and temporal scales:
  Local structure and the effect of host population size}},
\newblock \bibinfo{journal}{Science} \bibinfo{volume}{355}
  (\bibinfo{year}{2017}) \bibinfo{pages}{1302--1306}.
  \DOIprefix\doi{10.1126/science.aaj9384}.
%Type = Article
\bibitem[{Alam et~al.(2006)Alam, Hasan, Sadique, Bhuiyan, Ahmed, Nusrin, Nair,
  Siddique, Sack, Sack, Huq, and Colwell}]{Alam2006}
\bibinfo{author}{M.~Alam}, \bibinfo{author}{N.~A. Hasan},
  \bibinfo{author}{A.~Sadique}, \bibinfo{author}{N.~A. Bhuiyan},
  \bibinfo{author}{K.~U. Ahmed}, \bibinfo{author}{S.~Nusrin},
  \bibinfo{author}{G.~B. Nair}, \bibinfo{author}{A.~K. Siddique},
  \bibinfo{author}{R.~B. Sack}, \bibinfo{author}{D.~A. Sack},
  \bibinfo{author}{A.~Huq}, \bibinfo{author}{R.~R. Colwell},
\newblock \bibinfo{title}{{Seasonal Cholera Caused by Vibrio cholerae
  Serogroups O1 and O139 in the Coastal Aquatic Environment of Bangladesh}},
\newblock \bibinfo{journal}{Appl. Environ. Microbiol.} \bibinfo{volume}{72}
  (\bibinfo{year}{2006}) \bibinfo{pages}{4096--4104}.
  \DOIprefix\doi{10.1128/aem.00066-06}.
%Type = Book
\bibitem[{et~al Heymann(2008)}]{Heymann2008}
\bibinfo{author}{et~al Heymann}, \bibinfo{title}{{Control of Communicable
  Diseases Manual}}, \bibinfo{edition}{19th} ed., \bibinfo{publisher}{APHA},
  \bibinfo{year}{2008}. \DOIprefix\doi{10.1086/605668}.
%Type = Book
\bibitem[{Porta(2008)}]{Porta2008}
\bibinfo{author}{M.~Porta}, \bibinfo{title}{{A Dictionary of Epidemiology,
  Fifth Edition: Edited by Miquel Porta}}, \bibinfo{edition}{fifth} ed.,
  \bibinfo{publisher}{Oxford University Press}, \bibinfo{address}{New York},
  \bibinfo{year}{2008}.
%Type = Article
\bibitem[{Fraser(2007)}]{Fraser2007}
\bibinfo{author}{C.~Fraser},
\newblock \bibinfo{title}{{Estimating Individual and Household Reproduction
  Numbers in an Emerging Epidemic}},
\newblock \bibinfo{journal}{PLoS ONE}  (\bibinfo{year}{2007}).
  \DOIprefix\doi{10.1371/journal.pone.0000758}.
%Type = Misc
\bibitem[{Lessler(2018)}]{Lessler2018}
\bibinfo{author}{J.~Lessler}, \bibinfo{title}{{IDSpatialStats
  v0\textperiodcentered3\textperiodcentered7 R package read-only CRAN mirror}},
  \bibinfo{year}{2018}. \URLprefix
  \url{https://github.com/cran/IDSpatialStats}.
%Type = Article
\bibitem[{Grantz et~al.(2016)Grantz, Rane, Salje, Glass, Schachterle, and
  Cummings}]{Grantz2016}
\bibinfo{author}{K.~H. Grantz}, \bibinfo{author}{M.~S. Rane},
  \bibinfo{author}{H.~Salje}, \bibinfo{author}{G.~E. Glass},
  \bibinfo{author}{S.~E. Schachterle}, \bibinfo{author}{D.~A.~T. Cummings},
\newblock \bibinfo{title}{{Disparities in influenza mortality and transmission
  related to sociodemographic factors within Chicago in the pandemic of 1918}},
\newblock \bibinfo{journal}{PNAS} \bibinfo{volume}{113} (\bibinfo{year}{2016})
  \bibinfo{pages}{13839--13844}. \DOIprefix\doi{10.1073/pnas.1612838113}.
%Type = Article
\bibitem[{Grabowski et~al.(2014)Grabowski, Lessler, Redd, Kagaayi,
  Laeyendecker, Ndyanabo, Nelson, Cummings, Bwanika, Mueller, Reynolds,
  Munshaw, Ray, Lutalo, Manucci, Tobian, Chang, Beyrer, Jennings, Nalugoda,
  Serwadda, Wawer, Quinn, and Gray}]{Grabowski2014}
\bibinfo{author}{M.~K. Grabowski}, \bibinfo{author}{J.~Lessler},
  \bibinfo{author}{A.~D. Redd}, \bibinfo{author}{J.~Kagaayi},
  \bibinfo{author}{O.~Laeyendecker}, \bibinfo{author}{A.~Ndyanabo},
  \bibinfo{author}{M.~I. Nelson}, \bibinfo{author}{D.~A. Cummings},
  \bibinfo{author}{J.~B. Bwanika}, \bibinfo{author}{A.~C. Mueller},
  \bibinfo{author}{S.~J. Reynolds}, \bibinfo{author}{S.~Munshaw},
  \bibinfo{author}{S.~C. Ray}, \bibinfo{author}{T.~Lutalo},
  \bibinfo{author}{J.~Manucci}, \bibinfo{author}{A.~A. Tobian},
  \bibinfo{author}{L.~W. Chang}, \bibinfo{author}{C.~Beyrer},
  \bibinfo{author}{J.~M. Jennings}, \bibinfo{author}{F.~Nalugoda},
  \bibinfo{author}{D.~Serwadda}, \bibinfo{author}{M.~J. Wawer},
  \bibinfo{author}{T.~C. Quinn}, \bibinfo{author}{R.~H. Gray},
\newblock \bibinfo{title}{{The Role of Viral Introductions in Sustaining
  Community-Based HIV Epidemics in Rural Uganda: Evidence from Spatial
  Clustering, Phylogenetics, and Egocentric Transmission Models}},
\newblock \bibinfo{journal}{PLoS Med.} \bibinfo{volume}{11}
  (\bibinfo{year}{2014}). \DOIprefix\doi{10.1371/journal.pmed.1001610}.
%Type = Article
\bibitem[{Succo et~al.(2018)Succo, No{\"{e}}l, Nikolay, Maquart, Cochet,
  Leparc-Goffart, Catelinois, Salje, Pelat, de~Crouy-Chanel, de~Valk,
  Cauchemez, and Rousseau}]{Succo2018}
\bibinfo{author}{T.~Succo}, \bibinfo{author}{H.~No{\"{e}}l},
  \bibinfo{author}{B.~Nikolay}, \bibinfo{author}{M.~Maquart},
  \bibinfo{author}{A.~Cochet}, \bibinfo{author}{I.~Leparc-Goffart},
  \bibinfo{author}{O.~Catelinois}, \bibinfo{author}{H.~Salje},
  \bibinfo{author}{C.~Pelat}, \bibinfo{author}{P.~de~Crouy-Chanel},
  \bibinfo{author}{H.~de~Valk}, \bibinfo{author}{S.~Cauchemez},
  \bibinfo{author}{C.~Rousseau},
\newblock \bibinfo{title}{{Dengue serosurvey after a 2-month long outbreak in
  N{\^{i}}mes, France, 2015: was there more than met the eye?}},
\newblock \bibinfo{journal}{Eurosurveillance} \bibinfo{volume}{23}
  (\bibinfo{year}{2018}).
  \DOIprefix\doi{10.2807/1560-7917.ES.2018.23.23.1700482}.
%Type = Article
\bibitem[{Salje et~al.(2016)Salje, Cauchemez, Alera, Rodriguez-Barraquer,
  Thaisomboonsuk, Srikiatkhachorn, Lago, Villa, Klungthong, Tac-An, Fernandez,
  Velasco, {Roque Vito G.}, Nisalak, Macareo, Levy, Cummings, and
  Yoon}]{Salje2016}
\bibinfo{author}{H.~Salje}, \bibinfo{author}{S.~Cauchemez},
  \bibinfo{author}{M.~T. Alera}, \bibinfo{author}{I.~Rodriguez-Barraquer},
  \bibinfo{author}{B.~Thaisomboonsuk}, \bibinfo{author}{A.~Srikiatkhachorn},
  \bibinfo{author}{C.~B. Lago}, \bibinfo{author}{D.~Villa},
  \bibinfo{author}{C.~Klungthong}, \bibinfo{author}{I.~A. Tac-An},
  \bibinfo{author}{S.~Fernandez}, \bibinfo{author}{J.~M. Velasco},
  \bibinfo{author}{J.~{Roque Vito G.}}, \bibinfo{author}{A.~Nisalak},
  \bibinfo{author}{L.~R. Macareo}, \bibinfo{author}{J.~W. Levy},
  \bibinfo{author}{D.~Cummings}, \bibinfo{author}{I.-K. Yoon},
\newblock \bibinfo{title}{{Reconstruction of 60 Years of Chikungunya
  Epidemiology in the Philippines Demonstrates Episodic and Focal
  Transmission}},
\newblock \bibinfo{journal}{J. Inf. Dis.} \bibinfo{volume}{213}
  (\bibinfo{year}{2016}) \bibinfo{pages}{604--610}.
  \DOIprefix\doi{10.1093/infdis/jiv470}.
%Type = Article
\bibitem[{Bhoomiboonchoo et~al.(2014)Bhoomiboonchoo, Gibbons, Huang, Yoon,
  Buddhari, Nisalak, Chansatiporn, Thipayamongkolgul, Kalanarooj, Endy,
  Rothman, Srikiatkhachorn, Green, Mammen, Cummings, and
  Salje}]{Bhoomiboonchoo2014}
\bibinfo{author}{P.~Bhoomiboonchoo}, \bibinfo{author}{R.~V. Gibbons},
  \bibinfo{author}{A.~Huang}, \bibinfo{author}{I.-K. Yoon},
  \bibinfo{author}{D.~Buddhari}, \bibinfo{author}{A.~Nisalak},
  \bibinfo{author}{N.~Chansatiporn}, \bibinfo{author}{M.~Thipayamongkolgul},
  \bibinfo{author}{S.~Kalanarooj}, \bibinfo{author}{T.~Endy},
  \bibinfo{author}{A.~L. Rothman}, \bibinfo{author}{A.~Srikiatkhachorn},
  \bibinfo{author}{S.~Green}, \bibinfo{author}{M.~P. Mammen},
  \bibinfo{author}{D.~A. Cummings}, \bibinfo{author}{H.~Salje},
\newblock \bibinfo{title}{{The Spatial Dynamics of Dengue Virus in Kamphaeng
  Phet, Thailand}},
\newblock \bibinfo{journal}{PLoS Neglected Trop. Dis.} \bibinfo{volume}{8}
  (\bibinfo{year}{2014}) \bibinfo{pages}{6--11}.
  \DOIprefix\doi{10.1371/journal.pntd.0003138}.
%Type = Article
\bibitem[{Salje et~al.(2016)Salje, Lessler, Paul, Azman, Rahman, Rahman,
  Cummings, Gurley, and Cauchemez}]{Salje2016social}
\bibinfo{author}{H.~Salje}, \bibinfo{author}{J.~Lessler},
  \bibinfo{author}{K.~K. Paul}, \bibinfo{author}{A.~S. Azman},
  \bibinfo{author}{M.~W. Rahman}, \bibinfo{author}{M.~Rahman},
  \bibinfo{author}{D.~Cummings}, \bibinfo{author}{E.~S. Gurley},
  \bibinfo{author}{S.~Cauchemez},
\newblock \bibinfo{title}{{How social structures, space, and behaviors shape
  the spread of infectious diseases using chikungunya as a case study}},
\newblock \bibinfo{journal}{PNAS} \bibinfo{volume}{113} (\bibinfo{year}{2016})
  \bibinfo{pages}{13420--13425}. \DOIprefix\doi{10.1073/pnas.1611391113}.
%Type = Article
\bibitem[{Truelove et~al.(2019)Truelove, Graham, Moss, Metcalf, Ferrari, and
  Lessler}]{Truelove2019}
\bibinfo{author}{S.~A. Truelove}, \bibinfo{author}{M.~Graham},
  \bibinfo{author}{W.~J. Moss}, \bibinfo{author}{C.~J.~E. Metcalf},
  \bibinfo{author}{M.~J. Ferrari}, \bibinfo{author}{J.~Lessler},
\newblock \bibinfo{title}{{Characterizing the impact of spatial clustering of
  susceptibility for measles elimination}},
\newblock \bibinfo{journal}{Vaccine} \bibinfo{volume}{37}
  (\bibinfo{year}{2019}) \bibinfo{pages}{732--741}.
  \DOIprefix\doi{10.1016/j.vaccine.2018.12.012}.
%Type = Article
\bibitem[{Azman et~al.(2018)Azman, Luquero, Salje, Mba{\"{i}}bardoum, Adalbert,
  Ali, Bertuzzo, Finger, Toure, Massing, Ramazani, Saga, Allan, Olson, Leglise,
  Porten, and Lessler}]{Azman2018}
\bibinfo{author}{A.~S. Azman}, \bibinfo{author}{F.~J. Luquero},
  \bibinfo{author}{H.~Salje}, \bibinfo{author}{N.~N. Mba{\"{i}}bardoum},
  \bibinfo{author}{N.~Adalbert}, \bibinfo{author}{M.~Ali},
  \bibinfo{author}{E.~Bertuzzo}, \bibinfo{author}{F.~Finger},
  \bibinfo{author}{B.~Toure}, \bibinfo{author}{L.~A. Massing},
  \bibinfo{author}{R.~Ramazani}, \bibinfo{author}{B.~Saga},
  \bibinfo{author}{M.~Allan}, \bibinfo{author}{D.~Olson},
  \bibinfo{author}{J.~Leglise}, \bibinfo{author}{K.~Porten},
  \bibinfo{author}{J.~Lessler},
\newblock \bibinfo{title}{{Micro-Hotspots of Risk in Urban Cholera Epidemics}},
\newblock \bibinfo{journal}{J. Inf. Dis.} \bibinfo{volume}{218}
  (\bibinfo{year}{2018}) \bibinfo{pages}{1164--1168}.
  \DOIprefix\doi{10.1093/infdis/jiy283}.
%Type = Article
\bibitem[{Finger et~al.(2018)Finger, Bertuzzo, Luquero, Naibei, Tour{\'{e}},
  Allan, Porten, Lessler, Rinaldo, and Azman}]{Finger2018}
\bibinfo{author}{F.~Finger}, \bibinfo{author}{E.~Bertuzzo},
  \bibinfo{author}{F.~J. Luquero}, \bibinfo{author}{N.~Naibei},
  \bibinfo{author}{B.~Tour{\'{e}}}, \bibinfo{author}{M.~Allan},
  \bibinfo{author}{K.~Porten}, \bibinfo{author}{J.~Lessler},
  \bibinfo{author}{A.~Rinaldo}, \bibinfo{author}{A.~S. Azman},
\newblock \bibinfo{title}{{The potential impact of case-area targeted
  interventions in response to cholera outbreaks: A modeling study}},
\newblock \bibinfo{journal}{PLoS Med.} \bibinfo{volume}{15}
  (\bibinfo{year}{2018}) \bibinfo{pages}{1--27}.
  \DOIprefix\doi{10.1371/journal.pmed.1002509}.
%Type = Article
\bibitem[{{Hoang Quoc} et~al.(2016){Hoang Quoc}, Salje, Rodriguez-Barraquer,
  In-Kyu, Chau, Hung, Tuan, Lan, Willis, Nisalak, Kalayanarooj, Cummings, and
  Simmons}]{HoangQuoc2016}
\bibinfo{author}{C.~{Hoang Quoc}}, \bibinfo{author}{H.~Salje},
  \bibinfo{author}{I.~Rodriguez-Barraquer}, \bibinfo{author}{Y.~In-Kyu},
  \bibinfo{author}{N.~V.~V. Chau}, \bibinfo{author}{N.~T. Hung},
  \bibinfo{author}{H.~M. Tuan}, \bibinfo{author}{P.~T. Lan},
  \bibinfo{author}{B.~Willis}, \bibinfo{author}{A.~Nisalak},
  \bibinfo{author}{S.~Kalayanarooj}, \bibinfo{author}{D.~A. Cummings},
  \bibinfo{author}{C.~P. Simmons},
\newblock \bibinfo{title}{{Synchrony of Dengue Incidence in Ho Chi Minh City
  and Bangkok}},
\newblock \bibinfo{journal}{PLoS Neglected Trop. Dis.} \bibinfo{volume}{10}
  (\bibinfo{year}{2016}) \bibinfo{pages}{1--18}.
  \DOIprefix\doi{10.1371/journal.pntd.0005188}.
%Type = Article
\bibitem[{Salje et~al.(2018)Salje, Cummings, Rodriguez-Barraquer, Katzelnick,
  Lessler, Klungthong, Thaisomboonsuk, Nisalak, Weg, Ellison, Macareo, Yoon,
  Jarman, Thomas, Rothman, Endy, and Cauchemez}]{Salje2018}
\bibinfo{author}{H.~Salje}, \bibinfo{author}{D.~A.~T. Cummings},
  \bibinfo{author}{I.~Rodriguez-Barraquer}, \bibinfo{author}{L.~C. Katzelnick},
  \bibinfo{author}{J.~Lessler}, \bibinfo{author}{C.~Klungthong},
  \bibinfo{author}{B.~Thaisomboonsuk}, \bibinfo{author}{A.~Nisalak},
  \bibinfo{author}{A.~Weg}, \bibinfo{author}{D.~Ellison},
  \bibinfo{author}{L.~Macareo}, \bibinfo{author}{I.-K. Yoon},
  \bibinfo{author}{R.~Jarman}, \bibinfo{author}{S.~Thomas},
  \bibinfo{author}{A.~L. Rothman}, \bibinfo{author}{T.~Endy},
  \bibinfo{author}{S.~Cauchemez},
\newblock \bibinfo{title}{{Reconstruction of antibody dynamics and infection
  histories to evaluate dengue risk}},
\newblock \bibinfo{journal}{Nature} \bibinfo{volume}{557}
  (\bibinfo{year}{2018}) \bibinfo{pages}{719--723}.
  \DOIprefix\doi{10.1038/s41586-018-0157-4}.
%Type = Article
\bibitem[{Rehman et~al.(2018)Rehman, Salje, Kraemer, Subramanian, Cauchemez,
  Saif, and Chunara}]{Rehman2018}
\bibinfo{author}{N.~A. Rehman}, \bibinfo{author}{H.~Salje},
  \bibinfo{author}{M.~U.~G. Kraemer}, \bibinfo{author}{L.~Subramanian},
  \bibinfo{author}{S.~Cauchemez}, \bibinfo{author}{U.~Saif},
  \bibinfo{author}{R.~Chunara},
\newblock \bibinfo{title}{{Quantifying the impact of dengue containment
  activities using high-resolution observational data}},
\newblock \bibinfo{journal}{bioRxiv}  (\bibinfo{year}{2018}).
  \DOIprefix\doi{10.1101/401653}.
%Type = Article
\bibitem[{Levy et~al.(2015)Levy, Bhoomiboonchoo, Simasathien, Salje, Huang,
  Rangsin, Jarman, Fernandez, Klungthong, Hussem, Gibbons, and Yoon}]{Levy2015}
\bibinfo{author}{J.~W. Levy}, \bibinfo{author}{P.~Bhoomiboonchoo},
  \bibinfo{author}{S.~Simasathien}, \bibinfo{author}{H.~Salje},
  \bibinfo{author}{A.~Huang}, \bibinfo{author}{R.~Rangsin},
  \bibinfo{author}{R.~G. Jarman}, \bibinfo{author}{S.~Fernandez},
  \bibinfo{author}{C.~Klungthong}, \bibinfo{author}{K.~Hussem},
  \bibinfo{author}{R.~V. Gibbons}, \bibinfo{author}{I.-K. Yoon},
\newblock \bibinfo{title}{{Elevated transmission of upper respiratory illness
  among new recruits in military barracks in Thailand}},
\newblock \bibinfo{journal}{Influenza Respir. Viruses} \bibinfo{volume}{9}
  (\bibinfo{year}{2015}) \bibinfo{pages}{308--314}.
  \DOIprefix\doi{10.1111/irv.12345}.
%Type = Article
\bibitem[{Cummings(2009)}]{Cummings2009}
\bibinfo{author}{P.~Cummings},
\newblock \bibinfo{title}{{The Relative Merits of Risk Ratios and Odds
  Ratios}},
\newblock \bibinfo{journal}{JAMA Pediatr.} \bibinfo{volume}{163}
  (\bibinfo{year}{2009}) \bibinfo{pages}{438--445}.
  \DOIprefix\doi{10.1001/archpediatrics.2009.31}.
%Type = Article
\bibitem[{Boily et~al.(2009)Boily, Baggaley, Wang, Masse, White, Hayes, and
  Alary}]{Boily2009}
\bibinfo{author}{M.~C. Boily}, \bibinfo{author}{R.~F. Baggaley},
  \bibinfo{author}{L.~Wang}, \bibinfo{author}{B.~Masse}, \bibinfo{author}{R.~G.
  White}, \bibinfo{author}{R.~J. Hayes}, \bibinfo{author}{M.~Alary},
\newblock \bibinfo{title}{{Heterosexual risk of HIV-1 infection per sexual act:
  systematic review and meta-analysis of observational studies}},
\newblock \bibinfo{journal}{Lancet Inf. Dis.} \bibinfo{volume}{9}
  (\bibinfo{year}{2009}) \bibinfo{pages}{118--129}.
  \DOIprefix\doi{10.1016/S1473-3099(09)70021-0}.
%Type = Article
\bibitem[{Simpson and Mayer-Hasselwander(1986)}]{Simpson1986}
\bibinfo{author}{G.~Simpson}, \bibinfo{author}{Mayer-Hasselwander},
\newblock \bibinfo{title}{{Bootstrap sampling: applications in gamma-ray
  astronomy}},
\newblock \bibinfo{journal}{Astron. Astrophys.}  (\bibinfo{year}{1986})
  \bibinfo{pages}{340--348}.
%Type = Book
\bibitem[{Baddeley et~al.(2015)Baddeley, Rubak, and Turner}]{Baddeley2015}
\bibinfo{author}{A.~Baddeley}, \bibinfo{author}{E.~Rubak},
  \bibinfo{author}{R.~Turner}, \bibinfo{title}{{Spatial Point Patterns:
  Methodology and Applications with R}}, \bibinfo{edition}{first} ed.,
  \bibinfo{publisher}{CRC Press/Taylor {\&} Francis}, \bibinfo{address}{Boca
  Raton}, \bibinfo{year}{2015}. \DOIprefix\doi{10.1201/b19708}.
%Type = Article
\bibitem[{Reich et~al.(2011)Reich, Perl, Cummings, and Lessler}]{Reich2011}
\bibinfo{author}{N.~G. Reich}, \bibinfo{author}{T.~M. Perl},
  \bibinfo{author}{D.~A. Cummings}, \bibinfo{author}{J.~Lessler},
\newblock \bibinfo{title}{{Visualizing Clinical Evidence: Citation Networks for
  the Incubation Periods of Respiratory Viral Infections}},
\newblock \bibinfo{journal}{PLoS ONE} \bibinfo{volume}{6}
  (\bibinfo{year}{2011}). \DOIprefix\doi{10.1371/journal.pone.0019496}.
%Type = Article
\bibitem[{Weil et~al.(2009)Weil, Khan, Chowdhury, LaRocque, Faruque, Ryan,
  Calderwood, Qadri, and Harris}]{Weil2009}
\bibinfo{author}{A.~A. Weil}, \bibinfo{author}{A.~I. Khan},
  \bibinfo{author}{F.~Chowdhury}, \bibinfo{author}{R.~C. LaRocque},
  \bibinfo{author}{A.~S.~G. Faruque}, \bibinfo{author}{E.~T. Ryan},
  \bibinfo{author}{S.~B. Calderwood}, \bibinfo{author}{F.~Qadri},
  \bibinfo{author}{J.~B. Harris},
\newblock \bibinfo{title}{{Clinical Outcomes in Household Contacts of Patients
  with Cholera in Bangladesh}},
\newblock \bibinfo{journal}{Clin. Inf. Dis.} \bibinfo{volume}{49}
  (\bibinfo{year}{2009}) \bibinfo{pages}{1473--1479}.
  \DOIprefix\doi{10.1086/644779}.
%Type = Article
\bibitem[{Azman et~al.(2016)Azman, Rumunu, Abubakar, West, Ciglenecki,
  Helderman, Wamala, de~la Rosa~V{\'{a}}zquez, Perea, Sack, Legros, Martin,
  Lessler, and Luquero}]{Azman2016}
\bibinfo{author}{A.~S. Azman}, \bibinfo{author}{J.~Rumunu},
  \bibinfo{author}{A.~Abubakar}, \bibinfo{author}{H.~West},
  \bibinfo{author}{I.~Ciglenecki}, \bibinfo{author}{T.~Helderman},
  \bibinfo{author}{J.~F. Wamala}, \bibinfo{author}{R.~de~la
  Rosa~V{\'{a}}zquez}, \bibinfo{author}{W.~Perea}, \bibinfo{author}{D.~A.
  Sack}, \bibinfo{author}{D.~Legros}, \bibinfo{author}{S.~Martin},
  \bibinfo{author}{J.~Lessler}, \bibinfo{author}{F.~J. Luquero},
\newblock \bibinfo{title}{{Population-Level Effect of Cholera Vaccine on
  Displaced Populations, South Sudan, 2014}},
\newblock \bibinfo{journal}{Emerg. Infect. Dis.} \bibinfo{volume}{22}
  (\bibinfo{year}{2016}) \bibinfo{pages}{1067--1070}.
  \DOIprefix\doi{10.3201/eid2206.151592}.
%Type = Article
\bibitem[{Aldstadt et~al.(2012)Aldstadt, Yoon, Tannitisupawong, Jarman, Thomas,
  Gibbons, Uppapong, Iamsirithaworn, Rothman, Scott, and Endy}]{Aldstadt2012}
\bibinfo{author}{J.~Aldstadt}, \bibinfo{author}{I.-k. Yoon},
  \bibinfo{author}{D.~Tannitisupawong}, \bibinfo{author}{R.~G. Jarman},
  \bibinfo{author}{S.~J. Thomas}, \bibinfo{author}{R.~V. Gibbons},
  \bibinfo{author}{A.~Uppapong}, \bibinfo{author}{S.~Iamsirithaworn},
  \bibinfo{author}{A.~L. Rothman}, \bibinfo{author}{T.~W. Scott},
  \bibinfo{author}{T.~Endy},
\newblock \bibinfo{title}{{Space-time analysis of hospitalised dengue patients
  in rural Thailand reveals important temporal intervals in the pattern of
  dengue virus transmission}},
\newblock \bibinfo{journal}{Trop. Med. Int. Health} \bibinfo{volume}{17}
  (\bibinfo{year}{2012}) \bibinfo{pages}{1076--1085}.
  \DOIprefix\doi{10.1111/j.1365-3156.2012.03040.x}.
%Type = Article
\bibitem[{Vink et~al.(2014)Vink, Bootsma, and Wallinga}]{Vink2014}
\bibinfo{author}{M.~A. Vink}, \bibinfo{author}{M.~C.~J. Bootsma},
  \bibinfo{author}{J.~Wallinga},
\newblock \bibinfo{title}{{Serial Intervals of Respiratory Infectious Diseases:
  A Systematic Review and Analysis}},
\newblock \bibinfo{journal}{Am. J. Epidemiol.} \bibinfo{volume}{180}
  (\bibinfo{year}{2014}) \bibinfo{pages}{865--875}.
  \DOIprefix\doi{10.1093/aje/kwu209}.
%Type = Article
\bibitem[{Cori et~al.(2013)Cori, Ferguson, Fraser, and Cauchemez}]{Cori2013}
\bibinfo{author}{A.~Cori}, \bibinfo{author}{N.~M. Ferguson},
  \bibinfo{author}{C.~Fraser}, \bibinfo{author}{S.~Cauchemez},
\newblock \bibinfo{title}{{A New Framework and Software to Estimate
  Time-Varying Reproduction Numbers During Epidemics}},
\newblock \bibinfo{journal}{Am. J. Epidemiol.} \bibinfo{volume}{178}
  (\bibinfo{year}{2013}) \bibinfo{pages}{1505--1512}.
  \DOIprefix\doi{10.1093/aje/kwt133}.
%Type = Article
\bibitem[{Myllym{\"{a}}ki et~al.(2017)Myllym{\"{a}}ki, Myllym{\"{a}}ki, Tom,
  Mrkvi{\v{c}}ka, Grabarnik, Seijo, and Hahn}]{Myllymaki2017}
\bibinfo{author}{M.~Myllym{\"{a}}ki}, \bibinfo{author}{M.~Myllym{\"{a}}ki},
  \bibinfo{author}{T.~Tom}, \bibinfo{author}{T.~Mrkvi{\v{c}}ka},
  \bibinfo{author}{P.~Grabarnik}, \bibinfo{author}{H.~Seijo},
  \bibinfo{author}{U.~Hahn},
\newblock \bibinfo{title}{{Global envelope tests for spatial processes}},
\newblock \bibinfo{journal}{J. Royal Stat. Soc. Ser. B} \bibinfo{volume}{79}
  (\bibinfo{year}{2017}) \bibinfo{pages}{381--404}.
  \DOIprefix\doi{10.1111/rssb.12172}.
%Type = Article
\bibitem[{Salje et~al.(2016)Salje, Cummings, and Lessler}]{Salje2016Tx}
\bibinfo{author}{H.~Salje}, \bibinfo{author}{D.~A. Cummings},
  \bibinfo{author}{J.~Lessler},
\newblock \bibinfo{title}{{Estimating infectious disease transmission distances
  using the overall distribution of cases}},
\newblock \bibinfo{journal}{Epidemics} \bibinfo{volume}{17}
  (\bibinfo{year}{2016}) \bibinfo{pages}{10--18}.
  \DOIprefix\doi{10.1016/j.epidem.2016.10.001}.

\end{thebibliography}






% \end{thebibliography}

\end{document}